\documentclass[journal]{IEEEtran}

\usepackage{cite,amsmath,graphicx,subfigure,calc}

\usepackage{array} 

\newlength{\Lpr}
\newsavebox{\Bpr}

\newcommand{\V}[1]{\mbox{\boldmath$#1$\unboldmath}}

\newcommand{\diag}{\ensuremath{{\mathrm{diag}}}}

\newcommand{\re}{\ensuremath{{\mathrm{Re}}}}

\newcommand{\tr}{\ensuremath{{\mathrm{tr}}}}

\newcommand{\bdm}{\begin{displaymath}}
\newcommand{\edm}{\end{displaymath}}

\newcommand{\be}[1]{\begin{equation} \label{#1}}
\newcommand{\ee}{\end{equation}}

\newcommand{\bae}[3]{
\begin{equation} \label{#1}
\renewcommand{\arraystretch}{#2}
\begin{array}{#3}}

\newcommand{\eae}{\end{array}\end{equation}}

\newcommand{\baen}[2]{
\begin{displaymath} 
\renewcommand{\arraystretch}{#1}
\begin{array}{#2}}

\newcommand{\eaen}{\end{array}\end{displaymath}}

\newcommand{\DefLetter}[4]{
\newcommand{#1}{\ensuremath{\V{#2}}} 
\newcommand{#3}{\ensuremath{\V{#4}}} 
}

\DefLetter{\vzer}{0}{\mzer}{0}
\DefLetter{\vone}{1}{\mone}{1}
\DefLetter{\va}{a}{\ma}{A}
\DefLetter{\vb}{b}{\mb}{B}
\DefLetter{\vc}{c}{\mc}{C}
\DefLetter{\vd}{d}{\md}{D}
\DefLetter{\ve}{e}{\me}{E}
\DefLetter{\vf}{f}{\mf}{F}
\DefLetter{\vg}{g}{\mg}{G}
\DefLetter{\vh}{h}{\mh}{H}
\DefLetter{\vi}{i}{\mi}{I}
\DefLetter{\vj}{j}{\mj}{J}
\DefLetter{\vk}{k}{\mk}{K}
\DefLetter{\vl}{l}{\ml}{L}
\DefLetter{\vm}{m}{\mm}{M}
\DefLetter{\vn}{n}{\mn}{N}
\DefLetter{\vpr}{p}{\mpr}{P}
\DefLetter{\vq}{q}{\mq}{Q}
\DefLetter{\vr}{r}{\mr}{R}
\DefLetter{\vs}{s}{\ms}{S}
\DefLetter{\vt}{t}{\mt}{T}
\DefLetter{\vur}{u}{\mur}{U}
\DefLetter{\vv}{v}{\mv}{V}
\DefLetter{\vw}{w}{\mw}{W}
\DefLetter{\vx}{x}{\mx}{X}
\DefLetter{\vy}{y}{\my}{Y}
\DefLetter{\vz}{z}{\mz}{Z}

\DefLetter{\vdel}{\delta}{\mdel}{\Delta}
\DefLetter{\vphi}{\phi}{\mphi}{\Phi}
\DefLetter{\vpsi}{\psi}{\mpsi}{\Psi}
\DefLetter{\vrho}{\rho}{\mrho}{\Lambda}
\DefLetter{\vxi}{\xi}{\mxi}{\Xi}

\DefLetter{\valpha}{\alpha}{\malpha}{\Alpha}
\DefLetter{\vbeta}{\beta}{\mbeta}{\Beta}
\DefLetter{\vlam}{\lambda}{\mlam}{\Lambda}
\DefLetter{\vsig}{\sigma}{\msig}{\Sigma}
\DefLetter{\vtau}{\tau}{\mtau}{\tau}
\DefLetter{\vtheta}{\theta}{\mtheta}{\Theta}
\DefLetter{\vome}{\omega}{\mome}{\Omega}
\DefLetter{\vzero}{0}{\mzero}{0}
\DefLetter{\vgam}{\gamma}{\mgam}{\Gamma}
\DefLetter{\veps}{\epsilon}{\meps}{\Epsilon}
\DefLetter{\veta}{\eta}{\meta}{\Eta}

\newcommand{\DefFuncLetter}[2]{
\newcommand{#1}{\ensuremath{{#2}}} 
}

\DefFuncLetter{\Fzer}{0}
\DefFuncLetter{\Fa}{a}
\DefFuncLetter{\FA}{A}
\DefFuncLetter{\Fb}{b}
\DefFuncLetter{\Fc}{c}
\DefFuncLetter{\FC}{C}
\DefFuncLetter{\Fd}{d}
\DefFuncLetter{\Fe}{e}
\DefFuncLetter{\Ff}{f}
\DefFuncLetter{\Fg}{g}
\DefFuncLetter{\FG}{G}
\DefFuncLetter{\Fh}{h}
\DefFuncLetter{\FH}{H}
\DefFuncLetter{\Fi}{i}
\DefFuncLetter{\Fk}{k}
\DefFuncLetter{\Fl}{l}
\DefFuncLetter{\FL}{L}
\DefFuncLetter{\Fm}{m}
\DefFuncLetter{\Fn}{n}
\DefFuncLetter{\Fnr}{n}
\DefFuncLetter{\FN}{N}
\DefFuncLetter{\Fo}{o}
\DefFuncLetter{\FO}{O}
\DefFuncLetter{\Fpr}{p}
\DefFuncLetter{\FPr}{P}
\DefFuncLetter{\Fq}{q}
\DefFuncLetter{\Fr}{r}
\DefFuncLetter{\Fs}{s}
\DefFuncLetter{\FS}{S}
\DefFuncLetter{\Ft}{t}
\DefFuncLetter{\FT}{T}
\DefFuncLetter{\Fu}{u}
\DefFuncLetter{\FU}{U}
\DefFuncLetter{\Fv}{v}
\DefFuncLetter{\Fw}{w}
\DefFuncLetter{\FW}{W}
\DefFuncLetter{\Fx}{x}
\DefFuncLetter{\Fy}{y}
\DefFuncLetter{\FY}{Y}
\DefFuncLetter{\Fz}{z}
\DefFuncLetter{\FZ}{Z}
\DefFuncLetter{\Falp}{\alpha}
\DefFuncLetter{\Fbet}{\beta}
\DefFuncLetter{\Fchi}{\chi}
\DefFuncLetter{\Fdel}{\delta}
\DefFuncLetter{\Fzet}{\zeta}
\DefFuncLetter{\FEps}{\Epsilon}
\DefFuncLetter{\Feta}{\eta}
\DefFuncLetter{\Fphi}{\phi}
\DefFuncLetter{\FPhi}{\Phi}
\DefFuncLetter{\Fpsi}{\psi}
\DefFuncLetter{\FPsi}{\Psi}
\DefFuncLetter{\Fgam}{\gamma}
\DefFuncLetter{\FGam}{\Gamma}
\DefFuncLetter{\Flam}{\lambda}
\DefFuncLetter{\FLam}{\Lambda}
\DefFuncLetter{\Fsig}{\sigma}
\DefFuncLetter{\Ftau}{\tau}
\DefFuncLetter{\Fome}{\omega}
\DefFuncLetter{\Feps}{\epsilon}
\DefFuncLetter{\Fthe}{\theta}
\DefFuncLetter{\Fvar}{\vartheta}

\DefFuncLetter{\FB}{B}
\DefFuncLetter{\FD}{D}
\DefFuncLetter{\FE}{E}
\DefFuncLetter{\FF}{F}
\DefFuncLetter{\FI}{I}
\DefFuncLetter{\FJ}{J}
\DefFuncLetter{\FM}{M}
\DefFuncLetter{\FR}{R}
\DefFuncLetter{\FV}{V}
\DefFuncLetter{\FX}{X}

\newcommand{\DefCalLetter}[2]{
\newcommand{#1}{\ensuremath{\mathcal{#2}}} 
}
\DefCalLetter{\CC}{C}
\DefCalLetter{\CD}{D}

\DefCalLetter{\CS}{S}
\DefCalLetter{\CV}{V}

\newcommand{\DefSubLetter}[2]{
\newcommand{#1}{\mathrm{#2}} 
}

\DefSubLetter{\slzer}{0}
\DefSubLetter{\sla}{a}
\DefSubLetter{\slA}{A}
\DefSubLetter{\slb}{b}
\DefSubLetter{\slB}{B}
\DefSubLetter{\slc}{c}
\DefSubLetter{\slC}{C}
\DefSubLetter{\sld}{d}
\DefSubLetter{\slD}{D}
\DefSubLetter{\sle}{e}
\DefSubLetter{\slE}{E}
\DefSubLetter{\slf}{f}
\DefSubLetter{\slF}{F}
\DefSubLetter{\slg}{g}
\DefSubLetter{\slG}{G}
\DefSubLetter{\slh}{h}
\DefSubLetter{\slH}{H}
\DefSubLetter{\sli}{i}
\DefSubLetter{\slI}{I}
\DefSubLetter{\slk}{k}
\DefSubLetter{\sll}{l}
\DefSubLetter{\slL}{L}
\DefSubLetter{\slm}{m}
\DefSubLetter{\slM}{M}
\DefSubLetter{\sln}{n}
\DefSubLetter{\slnr}{n}
\DefSubLetter{\slN}{N}
\DefSubLetter{\slo}{o}
\DefSubLetter{\slp}{p}
\DefSubLetter{\slP}{P}
\DefSubLetter{\slq}{q}
\DefSubLetter{\slQ}{Q}
\DefSubLetter{\slr}{r}
\DefSubLetter{\slR}{R}
\DefSubLetter{\sls}{s}
\DefSubLetter{\slS}{S}
\DefSubLetter{\slt}{t}
\DefSubLetter{\slT}{T}
\DefSubLetter{\slu}{u}
\DefSubLetter{\slU}{U}
\DefSubLetter{\slv}{v}
\DefSubLetter{\slw}{w}
\DefSubLetter{\slW}{W}
\DefSubLetter{\slx}{x}
\DefSubLetter{\slX}{X}
\DefSubLetter{\sly}{y}
\DefSubLetter{\slY}{Y}
\DefSubLetter{\slz}{z}
\DefSubLetter{\slZ}{Z}

\DefSubLetter{\slalp}{\alpha}
\DefSubLetter{\slbet}{\beta}
\DefSubLetter{\sldel}{\delta}
\DefSubLetter{\slDel}{\Delta}
\DefSubLetter{\sleps}{\epsilon}
\DefSubLetter{\slgam}{\gamma}
\DefSubLetter{\slphi}{\phi}
\DefSubLetter{\sltau}{\tau}
\DefSubLetter{\slxi}{\xi}
\DefSubLetter{\slthe}{\theta}

\newcommand{\FigWide}[3]{
\begin{figure*}
\setlength{\unitlength}{1cm}
\centering{
\fbox{
\begin{picture}(17,5.5)
\thicklines
#1
\ifnum#3=1
\gridD
\fi
\end{picture}
}
}
\caption{#2}
\end{figure*}
}

\newcommand{\gridD}{
\linethickness{0.2mm}
\multiput(0,0)(0,1){11}{\line(1,0){10}}
\multiput(0,0)(1,0){11}{\line(0,1){10}}

\linethickness{0.05mm}
\multiput(0,0)(0,0.5){21}{\line(1,0){10}}
\multiput(0,0)(0.5,0){21}{\line(0,1){10}}

\linethickness{0.02mm}
\multiput(0,0)(0,0.1){101}{\line(1,0){10}}
\multiput(0,0)(0.1,0){101}{\line(0,1){10}}

\put(-0.09,-0.3){0} \put(0.91,-0.3){1} \put(1.91,-0.3){2} \
\put(2.91,-0.3){3} \put(3.91,-0.3){4} \put(4.91,-0.3){5} \
\put(5.91,-0.3){6} \put(6.91,-0.3){7} \put(7.91,-0.3){8} \
\put(8.91,-0.3){9} \put(9.91,-0.3){10}

\put(-0.25,-0.1){0} \put(-0.25,0.9){1} \put(-0.25,1.9){2} \
\put(-0.25,2.9){3} \put(-0.25,3.9){4} \put(-0.25,4.9){5} \
\put(-0.25,5.9){6} \put(-0.25,6.9){7} \put(-0.25,7.9){8} \
\put(-0.25,8.9){9} \put(-0.25,9.9){10}
}

\newcommand{\Fig}[4]{
\begin{figure}
\setlength{\unitlength}{1cm}
\centering{
\fbox{
\begin{picture}(8.5,#3)
\thicklines
#1
\ifnum#4=1
\gridD
\fi
\end{picture}
}
}
\caption{#2}
\end{figure}
}

\newcommand{\Sphio}{\mphi_o}
\newcommand{\Sphi}{\mphi}
\newcommand{\Srzd}{\mr_z}

\newcommand{\Srzl}{\mr_{z\lambda}}

\newcommand{\Srzlx}{\mr_{z\lambda,x}}
\newcommand{\Slam}{\mlam}
\newcommand{\Sld}{\FL_D}
\newcommand{\Sldo}{\FL_{Do}}
\newcommand{\Sls}{\FL_S}

\newcommand{\Smzl}{\mm_{z\lambda}}
\newcommand{\Smzlx}{\mm_{z\lambda,x}}
\newcommand{\Srs}{\mr_s}

\newcommand{\SiM}{\mi_M}
\newcommand{\SiK}{\mi_K}

\newcommand{\Sdphi}{\md}
\newcommand{\Sp}{\mpr}
\newcommand{\Spx}{\mpr_x}
\newcommand{\Spz}{\mpr_z}
\newcommand{\Spzx}{\mpr_{z,x}}

\newcommand{\Sre}{\textrm{Re}}
\newcommand{\krange}{k=1,\ldots,\,K}

\begin{document}
\title{Efficient ML Direction of Arrival Estimation assuming Unknown Sensor Noise Powers}

\author{J. Selva   
\thanks{Copyright (c) 2016 IEEE.  Personal use of this material is permitted. However, permission to use this material for any other purposes must be obtained from the IEEE by sending a request to pubs-permissions@ieee.org.  The author is with the Dept.  of Physics, Systems Engineering and Signal Theory (DFISTS), University of Alicante, P.O.Box 99, E-03080 Alicante, Spain (e-mail: jesus.selva@ua.es). This work has been supported by the Spanish Ministry of Economy and Competitiveness (MINECO) and EU FEDER under project TIN2014-55413-C2-2-P.}
}

\maketitle

\begin{abstract}

  This paper presents an efficient method for computing maximum likelihood (ML) direction of arrival (DOA) estimates assuming unknown sensor noise powers. The method combines efficient Alternate Projection (AP) procedures with Newton iterations. The efficiency of the method lies in the fact that all its intermediate steps have low complexity. The main contribution of this paper is the method's last step, in which a concentrated cost function is maximized in both the DOAs and noise powers in a few iterations through a Newton procedure. This step has low complexity because  it employs closed-form expressions of the cost function's gradients and Hessians, which are presented in the paper.  The method's total computational burden is of just a few mega-flops in typical cases. We present the method for the deterministic and stochastic ML estimators.  An analysis of the deterministic ML cost function's gradient reveals an unexpected drawback of its associated estimator: if the noise powers are unknown, then it is either degenerate or inconsistent. The root-mean-square (RMS) error performance and computational burden of the method are assessed numerically. 
  
\end{abstract}

\section{Introduction}

The estimation of the directions of arrival (DOA) of several narrowband sources is usually performed assuming uncorrelated noise, both spatially and temporally, and of the same power at each sensor. We may jointly term these three conditions the ``uniform noise assumption'' for short. This assumption greatly simplifies DOA estimation, because the noise can be described by a single parameter, its power. Besides, under this assumption, the array covariance matrix admits the well-known decomposition into a signal and a noise subspace, making it possible to employ estimators such as MUSIC (MUltiple SIgnal Classification) and ESPRIT (Estimation of Signal Parameters via Rotational Invariance Techniques).
In practice, however, this assumption may be inappropriate due to various causes, such as the presence of interference or imperfections in the sensors' processing chains.  Though the temporal noise correlation can be easily eliminated by working in the spectral domain, the spatial noise correlation and the unequal noise powers must be taken into account in the DOA estimation model.
General ways to model the noise when the uniform noise assumption fails were proposed in \cite{LeCadre89} and \cite {Friedlander95}, where the authors approximate the spatial noise covariance matrix using an  autoregressive model and a truncated Fourier series respectively. In both cases, the noise modeling adds new parameters that must also be estimated. 

An important class of signal models for DOA estimation drops the equal-noise-powers condition from the uniform noise assumption, i.e, in them the noise is viewed as uncorrelated, both spatially and temporally, but of unequal power at each sensor. These models assuming unequal noise powers (UNP) are important because they describe sparse sensors subject to different noise perturbations. 
The estimation of DOAs from these UNP models can be viewed as a more complex version of the usual uniform-noise DOA estimation, because it is necessary to cope with the noise powers as additional unknown parameters.  In this UNP case, the usual subspace methods, such as MUSIC and ESPRIT, are not directly usable and, besides, the ML estimators involve a more complex optimization problem. These difficulties have led to a variety of techniques that attempt to obtain DOA estimates with reasonable complexity under various assumptions. In \cite{Pesavento01}, the authors presented the deterministic and stochastic Cramer-Rao (CR) bounds, as well as the deterministic ML (DML) estimator for the UNP case.  In this work, the ML estimates were obtained through a genetic algorithm.  A related work was  \cite{Vorobyov05}, where a set of noise uncorrelated subarrays was considered. In \cite{Madurasinghe05} a simple estimator of the noise powers was proposed, provided there is a signal subspace estimate available. In \cite{Chen07}, the authors considered the same UNP model but for wideband signals and, in \cite{Chen08b}, the authors presented the stochastic ML (SML) estimator. In this last reference, the ML cost function was iteratively concentrated in the noise powers, signal parameters, and DOAs. The concentration in the DOAs was performed using the Alternate Projection (AP) method in \cite{Ziskind88}. Finally, the estimation of the signal subspace has been addressed in \cite{Liao16,Liao17}. 

The maximization of ML cost functions like those in the UNP scenario involve two basic operations that often appear mixed: one is the coarse localization of the global maximum, and the other is the refinement of that coarse localization in order to obtain the actual global maximum. Usually, these two operations are performed for increasing orders of the signal model, i.e, by successively adding DOAs. The total computational burden of such an optimization procedure is fundamentally given by the complexities of the coarse localization and refinement steps. Thus, one may attempt to locate the global maximum through a genetic algorithm, which is a high complexity procedure, or one may successively add DOAs using the Alternate Projection (AP) method in \cite{Ziskind88}, which just involves a linear search in each step. Also, one may perform the refinement through a coordinate descent method that involves a large number of one-dimensional optimizations, or one may use a Newton's method whose complexity can be very low, provided the computation of the cost function's gradient and Hessian is cheap. The most desirable case is that of a method based on low-complexity coarse localization and refinement steps. One such method has already been developed for the uniform-noise DOA problem in \cite{Selva05a,Selva17,Selva18b} and we may term it the Alternate Projection Newton (APN) method, given that it combines the line searches present in the AP method in \cite{Ziskind88} with Newton's method.  The APN method is efficient for two reasons. First, the line searches are efficiently implemented using interpolation techniques that include the use of the FFT. And second, the implementation of Newton's method has low complexity, because there exist compact expressions of the gradient and Hessian required in each Newton iteration.

In this paper, we present an extension of this APN method to the UNP case. The main contribution is an efficient computation method for Newton iterations that maximize a concentrated cost function in both the DOAs and sensor-noise powers. These iterations have low complexity because they make use of compact expressions of the cost function's gradient and Hessian.  Actually, as shown in the numerical examples, the total computational cost of computing the ML estimates is just of a few mega flops (MFlops), (Fig. \ref{fig:8}).  As a spin-off, the analysis of the gradient of the DML cost function will reveal  an unexpected drawback of the DML estimator: it is either degenerate or inconsistent. 

The paper has been organized as follows. In the next section, we introduce the UNP signal model and the DML and SML cost functions. Then, we introduce the APN method for both cost functions in Sec. \ref{sec:ap}, and comment on its efficient implementation in Sec. \ref{sec:ei}. Afterward, we present the gradient and Hessian expressions for the DML and SML cost functions in Secs. \ref{sec:ge} and \ref{sec:he}, which allow an efficient implementation of Newton's method. The expressions are formed by blocks corresponding to the angle and noise parameters. The derivations of these expressions have been omitted in the paper due to lack of space, except for two gradient blocks that are derived in Ap. \ref{sec:dg}, but they are available as complementary material.  In Sec. \ref{sec:ge}, we employ the gradient expression to check whether the DML and SML estimators are consistent asymptotically. It turns out the DML gradient is non-zero close to the true values of the angles of arrival and noise parameters at high signal-to-noise (SNR) ratios. The interpretation of this fact is that the estimator is either degenerate or inconsistent. In Sec. \ref{sec:ee}, we discuss the efficient evaluation of the gradient and Hessian expressions. Finally, we assess the APN method in Sec. \ref{sec:ne} numerically.

\subsection{Notations}

We will employ the following notations:

\begin{itemize}
\item Column vectors and matrices will be written in lower- and upper-case bold font respectively. Thus $\vx$ will denote a column vector and $\mx$ a matrix.
\item $[\mx]_{p,q}$ will denote the $(p,q)$ component of matrix $\mx$, and $[\mx]_{p,\cdot}$ and  $[\mx]_{\cdot,q}$ its $p$th row and $q$th column respectively.
\item $\diag(\vx)$ will be the diagonal matrix formed by the components of $\vx$.
\item For square $\mx$, $\tr\{\mx\}$ will denote the trace of $\mx$, i.e, the sum of its diagonal components.
  
\item $[\vx;a]$ will denote the column vector formed by appending the scalar $a$ to the column vector $\vx$.

\item $\mx^H$ will stand for the conjugate transpose of $\mx$.  
\item $\mx^\dagger$ will stand for the pseudo-inverse of matrix $\mx$. If $\mx$ has full column rank then $\mx^\dagger=(\mx^H\mx)^{-1}\mx^H$.

\item Given a matrix $\mx$ of full column rank, its projection matrix is $\mx \mx^\dagger$. 
\item The operator '$\equiv$' will indicate a symbol or function definition.
\item Given two variables $a$ and $b$, the arrow $a\rightarrow b$ will denote the replacement of $a$ with $b$ in a given expression. 

\item $\vdel_{K,p}$ will denote a $K\times 1$ Dirac vector
\begin{equation}
\label{eq:255}\nonumber
[\vdel_{K,p}]_k\equiv
\begin{cases}
  0, & k\neq p\\
  1, & k=p.
  \end{cases}
\end{equation}

\end{itemize}

Throughout the paper, we will often omit the dependency on the various parameters in writing for simplicity. Thus, for example, $\mphi$ will stand for $\mphi(\vtheta,\vlam)$ and $\mphi_o$ for $\mphi_o(\vtheta)$. The actual dependencies will be evident from the context. 

We will denote the various ML cost functions in the paper using the subscripts ``$o$'', ``D'', and ``S'',
\begin{itemize}
\item $L_{Do}(\vtheta)$ will denote the compressed deterministic ML cost function assuming uniform noise. 
  \item $L_D(\vtheta,\vlam)$ and $\FL_S(\vtheta,\vlam)$ will respectively denote the compressed deterministic and stochastic ML cost functions, assuming unknown noise powers. \end{itemize}

\section{Signal model and DML and SML cost functions}
\label{sec:sm}

We consider a linear array formed by $M$ sensors and $K$ waves impinging from angles of arrival $\theta_k$, $\krange$. If the receiver takes $N$ snapshots, the data model is
\begin{equation}
\label{eq:53}
\mz=\mphi_o(\vtheta)\ms+\mn,
\end{equation}
where
\begin{itemize}
\item $[\mz]_{m,n}$ is the $n$th sample from the $m$th sensor,
\item $\vtheta$ contains the $K$ angles of arrival (AOAs) $\theta_k$, $[\vtheta]_k\equiv \theta_k$,
\item $\vphi_o(\theta)$ is the array's response to a wave from angle $\theta$, 
\item $\mphi_o(\vtheta)$ is a matrix stacking the responses to the angles in $\vtheta$, $[\mphi_o(\vtheta)]_{\cdot,k}\equiv\vphi_o(\theta_k)$,

\item $[\ms]_{k,n}$ is the $n$th sample from the $k$th impinging signal,
 
\item and all components $[\mn]_{m,n}$ are independent noise samples that follow a complex normal circularly-symmetric distribution of zero mean and deviation $1/\lambda_m$, where we refer to $\lambda_m$ as the inverse noise deviation at the $m$th sensor.

\end{itemize}
Additionally, we  define the following vector and diagonal matrix from $\lambda_m$,
\begin{equation}
\label{eq:99a}\nonumber
[\vlam]_m\equiv \lambda_m,\;\; \mlam\equiv\diag(\vlam).
\end{equation}
Using $\mlam$, we have that the columns of $\mn$ have covariance  matrix $\mlam^{-2}$.

Let us now derive the compressed deterministic and stochastic ML cost functions for this model. To do so, it is convenient to recall first the complex Gaussian probability density function (PDF), \cite[Th. 15.1]{Kay93}. If the expected value of $\mz$ is the matrix $\me_z$ and the columns of $\mz$ are independent and have equal covariance matrix $\mc_z$, then the PDF of $\mz$ is
\begin{equation}
  \label{eq:105}
  \begin{split}
    \Ff(\mz)=&\frac{1}{\pi^{MN}|\mc_z|^N}\cdot\\
    &{}\;\;\;\;\exp(-\tr\{\mc_z^{-1}(\mz-\me_z)(\mz-\me_z)^H\}).
\end{split}
\end{equation}
If $\ms$ is viewed as a deterministic matrix, then $\mz$ in (\ref{eq:53}) has mean $\me_z\rightarrow \mphi_o\ms$ and covariance $\mc_z\rightarrow \mlam^{-2}$ and, from (\ref{eq:105}), the PDF of $\mz$ is
\begin{multline}
\label{eq:104}
\Ff_{D}(\mz|\vtheta,\vlam,\ms)\equiv\\
\frac{|\mlam|^{2N}}{\pi^{MN}}
\exp\Big({-\tr\Big\{\mlam^2(\mz-\mphi_o\ms)(\mz-\mphi_o\ms)^H\Big\}}\Big),
\end{multline}
where we have written $\mphi_o$ rather than $\mphi_o(\vtheta)$ for simplicity. If we take the logarithm of this expression and introduce the following ``whitened'' signature matrix
\begin{equation}
\label{eq:57}
\mphi(\vtheta,\vlam)\equiv\mlam\mphi_o(\vtheta),
\end{equation}
then, after straightforward manipulations, we obtain from (\ref{eq:104}) the cost function
\begin{equation}
  \label{eq:19}
  \begin{split}
    \FL_{D}(\vtheta,\vlam,\ms)\equiv & -NM\log\pi+2N\log|\mlam|\\
    &-\textrm{tr}\{(\mlam\mz-\mphi\ms)(\mlam\mz-\mphi\ms)^H\}).
\end{split}
\end{equation}
Next, as is well known, this expression is maximized in $\ms$ for fixed $\vtheta$ and $\vlam$, if the product $\mphi\ms$ is replaced with $\mpr\mlam\mz$, where $\mpr$ is the projection matrix of $\mphi$. For later use, we express this last matrix as
\begin{equation}
\label{eq:66}\nonumber
\mpr\equiv \mphi\mm\mphi^H,
\end{equation}
where $\mm$ is the inverse correlation matrix of $\mphi$,
\begin{equation}
\label{eq:61}\nonumber
\mm\equiv(\mphi^H\mphi)^{-1}.
\end{equation}
So if we replace $\mphi\ms$ with $\mpr\mlam\mz$ in (\ref{eq:19}) and perform straight-forward manipulations, we obtain the new cost function
\begin{equation}
\label{eq:59}
\FL_{D}(\vtheta,\vlam)\equiv  N\big(2\log|\mlam|-\textrm{tr}\{(\mi_M-\mpr)\mr_{z\lambda}\}
\big),
\end{equation}
where we have neglected the constant $-NM\log\pi$ in (\ref{eq:19}) and $\mr_{z\lambda}$ denotes the ``whitened'' data correlation matrix
\begin{equation}
\label{eq:90}
\mr_{z\lambda}\equiv\frac{1}{N}\mlam\mz\mz^H\mlam.
\end{equation}
(\ref{eq:59}) is the compressed DML cost function that will be used in the rest of the paper.

For uniform noise, the cost function equivalent to $\FL_D$ can be easily derived from (\ref{eq:59}), simply by setting $\mlam=\SiM$. (For a proof see \cite[Sec. 4.4.2]{Haykin93}.) Since this function will be instrumental in the paper, it is convenient to introduce it now. The uniform-noise DML cost function is 
\begin{equation}
\label{eq:91}
\FL_{Do}(\vtheta)\equiv -N\textrm{tr}\{(\mi_M-\mpr_o)\mr_{z}\}
\big),
\end{equation}
where the sub-script ``$o$'' indicates that the matrices are computed with $\mlam=\SiM$ and
\begin{equation}
\label{eq:96}\nonumber
\mr_z\equiv \frac{1}{N}\mz\mz^H.
\end{equation}

Next, let us introduce the stochastic ML cost function. In the stochastic modeling, the columns of $\ms$ are viewed as independent trials of a complex Gaussian distribution of zero mean and covariance $\mr_s$ and, from (\ref{eq:53}), this leads to the PDF in (\ref{eq:105}) with $\me_z\rightarrow \mzer$ and $\mc_z\rightarrow\mphi_o\mr_s\mphi_o^H+\mlam^{-2}$. For simplicity, let us write this last covariance matrix as
\begin{equation}
\label{eq:106}\nonumber
\mphi_o\mr_s\mphi_o^H+\mlam^{-2}=\mlam^{-1}(\mphi\mr_s\mphi^H+\SiM)\mlam^{-1}.
\end{equation}
Substituting these values of $\me_z$ and $\mc_z$ into (\ref{eq:105}), we obtain the PDF in the stochastic case
\begin{multline}
\label{eq:64}
\Ff_{S}(\mz|\vtheta,\vlam,\mr_s)\equiv
\frac{|\mlam|^{2N}}{\pi^{MN}|\mi_M+\mphi\mr_s\mphi^H|^N}\cdot\\
\exp\Big({-N\tr\Big\{(\mi_M+\mphi\mr_s\mphi^H)^{-1}\mr_{z\lambda}\Big\}}\Big),
\end{multline}
where we have inserted the signature matrix in (\ref{eq:57}) and the whitened correlation matrix in (\ref{eq:90}).

Taking the logarithm of (\ref{eq:64}), we obtain the cost function 
\begin{equation}
  \label{eq:21}
  \begin{split}
\FL_{S}(\vtheta,\vlam,\mr_s)&\equiv
-MN\log(\pi)+2N\log|\mlam|  \\ -N\log|\mi_M&+\mphi\mr_s\mphi^H|
-N\textrm{tr}
\{(\mi_M+\mphi\mr_s\mphi^H)^{-1}\mr_{z\lambda}\}.
\end{split}
\end{equation}
This expression can be maximized in $\mr_s$ for fixed $\vtheta$ and $\vlam$ and the maximum is attained at
\begin{equation}
\label{eq:67}
\hat\mr_s\equiv \mphi^\dagger\Srzl(\mphi^\dagger)^H-\mm.
\end{equation}
(See \cite{Jaffer88} for a proof.)

In order to replace $\mr_s$ with $\hat\mr_s$ in (\ref{eq:21}), it is convenient to start by performing this same replacement on the  covariance matrix appearing twice in (\ref{eq:21}), namely the matrix
\begin{equation}
\label{eq:120}
\mc\equiv\SiM+\mphi\hat\mr_s\mphi^H.
\end{equation}
More precisely, we proceed to derive compact expressions of $\mc$ and $\mc^{-1}$ in terms of $\mphi$ and $\Sp$.

First, noting that ${\mphi\mm\mphi^H=\mpr}$ and ${\mphi\mphi^\dagger=\mpr}$, the substitution of (\ref{eq:67}) into (\ref{eq:120}) yields the desired expression for $\mc$,
\begin{equation}
\label{eq:109}
\begin{split}
  \mc & = \SiM+\mphi(\mphi^\dagger\Srzl(\mphi^\dagger)^H-\mm)\mphi^H\\
  & = \SiM-\mpr+\mpr\Srzl\mpr.
\end{split}
\end{equation}

Second, regarding  $\mc^{-1}$, consider the QR decomposition ${\mphi=}{\mq\mr}$, with $\mq^H\mq=\mi_K$ and invertible $\mr$, and an $M\times (M-K)$ matrix $\mq_\perp$ spanning the orthogonal complement to $\mq$, ($\mq_\perp^H\mq_\perp=\mi_{M-K}$, $\mq^H\mq_\perp=\mzer$). Noting that $\mpr=\mq\mq^H$ and $\SiM-\mpr=\mq_\perp\mq_\perp^H$, we may write $\mc$ in (\ref{eq:109}) as
\begin{equation}
\label{eq:121}\nonumber
\mc =
\begin{bmatrix}
  \mq_\perp,\,\mq
\end{bmatrix}
\begin{bmatrix}
  \mi_{M-K}&\mzer\\
  \mzer &\mq^H\Srzl\mq
\end{bmatrix}
\begin{bmatrix}
 \mq_\perp^H\\ \mq^H
\end{bmatrix}.
\end{equation}
From this factorization, its clear that $\mc^{-1}$ is given by
\begin{equation}
\label{eq:122}\nonumber
\mc^{-1} =
\begin{bmatrix}
  \mq_\perp,\,\mq
\end{bmatrix}
\begin{bmatrix}
  \mi_{M-K}&\mzer\\
  \mzer &(\mq^H\Srzl\mq)^{-1}
\end{bmatrix}
\begin{bmatrix}
 \mq_\perp^H\\ \mq^H
\end{bmatrix}.
\end{equation}
And, finally, replacing  $\mq \rightarrow \mphi\mr^{-1}$, we obtain an expression for $\mc^{-1}$ in terms of $\mphi$ and $\mpr$ only:
\begin{equation}
\label{eq:124}\nonumber
\begin{split}
\mc^{-1}&=\mq_\perp\mq_\perp^H+\mq(\mq^H\Srzl\mq)^{-1}\mq^H\\
&=\SiM-\mpr\\
& {}\;\;\;\;+\mphi\mr^{-1}((\mr^{-1})^H\mphi^H\Srzl\mphi\mr^{-1})^{-1}(\mr^{-1})^H
\mphi^H\\
&=\SiM-\mpr+\mphi(\mphi^H\Srzl\mphi)^{-1}\mphi^H.
\end{split}
\end{equation}
We may write this formula concisely as
\begin{equation}
\label{eq:125}
\mc^{-1}=\SiM-\Sp+\Spz,
\end{equation}
where
\begin{equation}
\label{eq:112}
  \Spz\equiv\Sphi\Smzl\Sphi^H\;\textrm{and}\,\, \Smzl\equiv (\Sphi^H\Srzl\Sphi)^{-1}.
\end{equation}

Coming back to (\ref{eq:21}), the replacement of $\mr_s$ with $\hat\mr_s$ can be effected by substituting into that equation the identities (\ref{eq:109}), (\ref{eq:125}), and 
\begin{equation}
\label{eq:113}\nonumber
\tr\{\Spz\Srzl\}=K.
\end{equation}
This last identity can be easily deduced from (\ref{eq:112}). The result of these substitutions, neglecting constant summands, is the compressed cost function
\begin{equation}
\label{eq:68}
 \FL_{S}(\vtheta,\vlam)\equiv  N\big(2\log|\mlam|-\textrm{tr}\{(\mi_M-\mpr)\mr_{z\lambda}\}
 -\log|\mc|\big).
\end{equation}
This is the compressed stochastic ML cost function that will be analyzed in the rest of the paper. Note that $\FL_{S}$ in (\ref{eq:68}) is formed by adding a single term to $\FL_{D}$ in (\ref{eq:59}). Actually, we have

\begin{equation}
\label{eq:23}
\FL_{S}=\FL_{D}+\FL_{C},\;\text{where}\;\;\FL_{C}\equiv -N\log|\mc|.
\end{equation}

\section{The Alternate Projection Newton (APN) method}
\label{sec:ap}

The APN method for $\Sld$ and $\Sls$ is an extension of the method with the same name for $\Sldo$ and consists of three steps. In the first, we apply the APN method to $\FL_{Do} ( \vtheta )$ in order to obtain an initial estimate of $\vtheta$. Then, we apply the covariance matrix fitting method in \cite{Madurasinghe05} to obtain an initial estimate of $\vlam$. And finally, we refine these initial estimates of $\vtheta$ and $\vlam$ in order to obtain either the DML or SML estimates, by applying Newton's method to the corresponding cost function (either $\Sld$ or $\Sls$). We explain these three steps in the next sub-sections for $\Sls$. For $\Sld$ the steps would be analogous, but we will show in Sec. \ref{sec:ge} that there is a drawback in employing $\Sld$.

\subsection{APN method for the uniform-noise cost function $\FL_{Do}$}

The APN method for $\FL_{Do}$ looks for its global maximum by sequentially constructing vectors of $k$ angle estimates $\vtheta_k$, $k=0,\,1,\ldots,\,K$. Given $\vtheta_k$, the next vector $\vtheta_{k+1}$ is constructed by means of the following two sub-steps,  
\begin{itemize}
\item {\bf Add angle.} The method finds out the maximum of the function $\FL_{Do}([\vtheta_k;\theta])$ for varying $\theta$ and fixed $\vtheta_k$, and appends the corresponding abscissa to $\vtheta_k$ in order to form a new vector $\vtheta_{k+1,0}$. The initial set of estimates is the empty vector $\vtheta_{0}$.

\item {\bf Refinement.} $\vtheta_{k+1,0}$ is improved through a Newton iteration, ($r=0,\,1,\ldots$),
\begin{equation}
\label{eq:86}
\vtheta_{k,r+1}=\vtheta_{k,r}-\mu_{k,r}\hat\mh_{Do}^{-1}(\vtheta_{k,r})\,\vg_{Do}(\vtheta_{k,r}),
\end{equation}
where

\begin{itemize}
\item $\mu_{k,r}$ is initially equal to one, but can be reduced to a value ${0<\mu_{k,r}<1}$ in order to ensure that $\FL_{Do}(\vtheta_{k,r+1})>\FL_{Do}(\vtheta_{k,r})$.
  (See \cite[Ch. 6]{Dennis96} for the selection of $\mu_{k,r}$.)
  
\item $\hat\mh_{Do}(\vtheta_{k,r})$ is the Hessian  of $\FL_{Do}(\vtheta_{k,r})$ or an approximation to this matrix, but possibly perturbed to ensure that it is negative definite. This perturbation can be performed efficiently through a special Cholesky decomposition \cite[Sec. A5.5.2]{Dennis96}.
  
\item $\vg_{Do}(\vtheta_{k,r})$ is the gradient of $\FL_{Do}(\vtheta_{k,r})$.
  \end{itemize}

(\ref{eq:86}) is repeated until $\|\vtheta_{k,r+1}-\vtheta_{k,r}\|$ is sufficiently small. Then, the final vector $\vtheta_{k,r+1}$ is the new vector of angle estimates $\vtheta_{k+1}$.
\end{itemize}

\subsection{Initial noise parameter estimates}

In the second step, we denote $\vtheta'_0$ to the first step's output and compute an initial estimate of $\vlam$ using the method in \cite{Madurasinghe05}. Specifically, if $\vtheta'_0$ is close to the true value of $\vtheta$, then we may expect
\begin{equation}
\label{eq:115}\nonumber
\Srzd\approx \mphi_o(\vtheta_0')\ms\ms^H\mphi_o(\vtheta_0')^H+\mlam^{-2}
\end{equation}
and, therefore, the columns of $\Srzd-\mlam^{-2}$ approximately lie in the span of $\mphi_o(\vtheta_0')$. This implies that their projection onto the orthogonal complement to this last span is approximately zero, i.e, 
\begin{equation}
\label{eq:116}\nonumber
(\SiM-\Sp_o(\vtheta_0'))(\Srzd-\mlam^{-2})\approx \mzer.
\end{equation}
Thus, we may estimate $\vlam$ as the vector minimizing the Frobenius norm of this last matrix, given by
\begin{equation}
\label{eq:117}\nonumber
\tr\big\{(\SiM-\Sp_o(\vtheta_0'))(\Srzd-\mlam^{-2})(\Srzd-\mlam^{-2})^H\big\}.
\end{equation}
As can be readily checked \cite{Madurasinghe05}, the resulting estimate of $\vlam$ is
%
%
\begin{equation}
\label{eq:118}
[\vlam'_0]_m=\sqrt{\frac{1-[\mpr_o(\vtheta_0')]_{m,m}}
  {[\Srzd(\SiM-\mpr_o(\vtheta'_0))]_{m,m}}}.
\end{equation}

\subsection{Newton refinement of SML cost function $\FL_S$}

In this final step, we refine the initial estimate $[\vtheta_0';\vlam'_0]$ through Newton's method using an iteration similar to (\ref{eq:86}), but for $\FL_S(\vtheta,\vlam)$,
\begin{equation}
\label{eq:87}
\left[\begin{smallmatrix}
\vtheta'_{r+1}\\
\vlam'_{r+1}
\end{smallmatrix}
\right]
=
\left[\begin{smallmatrix}
\vtheta'_{r}\\
\vlam'_{r}
\end{smallmatrix}
\right]
-\mu_{r}\hat\mh_{S}^{-1}(\vtheta'_{r},\vlam_{r}')\,
\vg_{S}(\vtheta'_{r},\vlam_{r}').
\end{equation}

\section{Efficient implementation of the APN method -- Derivation of compact gradient and Hessian expressions}
\label{sec:ei}

The APN method in the previous section involves three kinds of operations: line searches on $\FL_{Do}$, evaluation of the formula in (\ref{eq:118}),  and the application of Newton's method to either $\FL_{Do}$ or $\FL_S$. The complexity of the line searches is reasonably small, given that only a coarse estimate of the maximum abscissa is required. Besides, this coarse estimate can be refined using a  proper interpolation method, \cite[Sec. IV]{Selva18b}. The evaluation of (\ref{eq:118}) is a simple, low-complexity operation. Finally, as to Newton's method, it is well known that it converges in a small number of iterations \cite{Dennis96} (quadratically, if sufficiently close to a maximum). Therefore, its complexity is mainly given by the computation of the cost function's value, gradient, and Hessian (or approximate Hessian).  For $\Sldo$, this last computation  is cheap, because there exist compact expressions for $\FL_{Do}$, $\vg_{Do}$ and $\hat\mh_{Do}$, which are  (\ref{eq:91}) and the following two expressions:
\begin{equation}
\label{eq:88}\nonumber
\vg_{Do}\equiv 2N\textrm{Re}\{\textrm{diag}\{\mphi_o^\dagger\mr_{z}(\mi_M-\
\mpr_o)\md_o\}\}
\end{equation}
and
\begin{equation}
\label{eq:89}  
\begin{split}
\hat\mh_{Do}\equiv   -2N\textrm{Re}\Big\{& \\  
(\mphi_o^\dagger\mr_{z}(\mphi_o^\dagger)&^H)\odot (\md_o^H(\mi_M-\mpr_o)\md_o)^T\Big\}.
\end{split}
\end{equation}
Actually, $\FL_{Do}$, $\vg_{Do}$ and  $\hat\mh_{Do}$ can be jointly computed in a small number of operations using the QR decomposition \cite[Sec. 4.6.4b]{Haykin93}.

The only missing aspect of the APN method is the computation of the gradient and Hessian of $\Sld$ and $\Sls$. At this point, the surprising fact is that there exist compact expressions for these differentials that allow their efficient computation, and this is also so for other cost functions in array processing such as $\Sldo$. The reason for this lies in the special structure of these cost functions, that we may summarize in the following two properties,
\begin{enumerate}
\item Each of these functions consists of summands of either the form $\tr\{\mf(\vx)\}$) or $\log|\mf(\vx)|$, where $\vx$ is the vector of variables and $\mf$ a matrix-valued function. 
\item $\mf(\vx)$ depends on the variables through matrices of the form $\mgam(\vx_1)$, where $\vx_1$ contains a subset of the variables in $\vx$. Besides, the  number of columns of $\mgam(\vx)$ is equal to the number of components of $\vx_1$, and the $k$th column of $\mgam(\vx_1)$ depends exclusively on the $k$th component of $\vx_1$, $[\vx_1]_k$.
\end{enumerate}

The way these properties facilitate the computation of compact gradient and Hessian expressions is better understood in a specific case. Thus, for example, consider the cost function $\FL_{Do}$ and assume we required to compute its Hessian. Recalling (\ref{eq:91}), we can readily see that $\Sldo$ depends on $\vtheta$ exclusively through $\Sphio$ and that this last matrix fulfills 2) above. This column-wise dependency of $\Sphio$ on $\vtheta$ allows us to differentiate $\Sphio$ once or twice through the handy formulas
\begin{gather}
\label{eq:78}
\frac{\partial}{\partial\theta_p}\mphi_o=\md_o\vdel_{K,p}\vdel_{K,p}^T,\;\;\;\;\\
\label{eq:258}
\frac{\partial^2}{\partial\theta_q\partial\theta_p}\mphi_o=
\md_{o2}\vdel_{K,p}\vdel_{K,p}^T\vdel_{K,q}\vdel_{K,q}^T,
\end{gather}
where
\begin{equation}
\label{eq:80}\nonumber
[\md_o]_{\cdot,k}\equiv \frac{\partial}{\partial\theta_k}\vphi_o(\theta_k),\;\;
[\md_{o2}]_{\cdot,k}\equiv \frac{\partial^2}{\partial\theta_k^2}\vphi_o(\theta_k).
\end{equation}

Next, let us consider the Hessian. Its $(p,q)$ component is 
\begin{equation}
\label{eq:77}
[\mh_{Do}]_{p,q}=\frac{\partial^2}{\partial{\theta_p}\partial{\theta_q}}\FL_{Do},
\end{equation}
and
we would need to compute $K(K+1)/2$ double differentials like this one in order to obtain the full matrix $\mh_{Do}$. However, by exploiting (\ref{eq:78}), (\ref{eq:258}), and the properties of the trace operator [property 1) above], we may obtain $\mh_{Do}$ from just one such double differential as follows.

First, substitute (\ref{eq:91}) into (\ref{eq:77}) and switch the trace and double derivative operators,
\begin{equation}
\label{eq:85}\nonumber
[\mh_{Do}]_{p,q}=-N\textrm{tr}\big\{\frac{\partial^2}{\partial{\theta_p}\partial{\theta_q}}\big((\mi_M-\mpr_o)\mr_{z}\big)\big\}.
\end{equation}

Second, apply the usual differentiation rules (including the rule for the matrix inverse) but resorting to (\ref{eq:78}) and (\ref{eq:258}) whenever $\mphi_o$ is encountered. Additionally, apply the property $\tr\{\ma\mb\}=\tr\{\mb\ma\}$ whenever necessary in order to place $\vdel_{K,q}^T$ on the left side inside any trace terms. This step is laborious but, as can be readily checked, it produces a sum of, say, $R$ terms of the form
\begin{equation}
\label{eq:83}\nonumber
[\mh_{Do}]_{p,q}=-N\sum_{r=1}^R\textrm{tr}\{\vdel_{K,q}^T\ma_r\vdel_{K,p}\vdel_{K,p}^T
\mb_r\vdel_{K,q}\},
\end{equation}
for specific matrices $\ma_r$ and $\mb_r$.

Third, we simplify the trace term in this expression since it is equal to $[\ma_r]_{q,p}[\mb_r]_{p,q}$, i.e,
\begin{equation}
\label{eq:259}\nonumber
[\mh_{Do}]_{p,q}=-N\sum_{r=1}^R[\ma_r]_{q,p}[\mb_r]_{p,q}.
\end{equation}

And fourth, since the matrices $\ma_r$ and $\mb_r$ are independent of $p$ and $q$, we may finally deduce that the Hessian is 
\begin{equation}
\label{eq:84}\nonumber
\mh_{Do}=-N\sum_{r=1}^R\mb_r\odot\ma_r^T.
\end{equation}

The analytical procedure specified by the last four steps has been used in the literature to obtain the exact Hessian of (\ref{eq:68}) in \cite[Sec. 5.5.2]{Selva04c}, which is given by the formula
\begin{equation}
\label{eq:244}
{\renewcommand{\arraystretch}{1.5}
\begin{array}{r@{\,}l} 
\mh_{Do}
&=2N\Sre\Big\{ \mm_o\odot(\md_o^H
    (\SiM-\mpr)\mr_{z}(\SiM-\mpr)\md_o)^T
  \\
  &{}\hspace{0.5cm}-
(\Sphi_o^\dagger \md_o)\odot(\Sphi_o^\dagger\mr_{z}(\SiM-\mpr)\md_o)^T
        \\
  &{}\hspace{0.5cm} -
(\Sphi_o^\dagger\mr_{z}(\SiM-\mpr)\md_o)\odot(\Sphi_o^\dagger\md_o)^T 
  \\
  & {}\hspace{0.5cm}-
(\Sphi_o^\dagger\mr_{z}(\Sphi_o^\dagger)^H)\odot(\md_o^H(\SiM-\mpr_o)\md_o)^T
    \\
  &{} \hspace{0.5cm}+
\SiK\odot(\Sphi_o^\dagger
   \mr_{z}(\SiM-\mpr)\md_{o2})^T
   \Big\},
\end{array}}
\end{equation}
where the subscript ``o'' means that the corresponding matrix has been computed for $\mlam=\SiM$.

Now, let us consider $\FL_D$ and $\FL_S$. These functions also fulfill the two properties above, where the matrices with property 2)  are $\mphi_o$ and $\mlam$; i.e, the $k$th column of $\mphi_o$ and the $m$th column of $\mlam$ depend exclusively on $\theta_k$ and $\lambda_m$ respectively, and both $\FL_D$ and $\FL_S$ only depend on any variable through $\mphi_o$ or $\mlam$. This implies that the procedure we have just described for $\FL_{Do}$ is also applicable to $\FL_D$ and $\FL_S$. We present in the next two sections the formulas that result from such application. First, we introduce the gradients of $\FL_D$ and $\FL_S$ in the next section and, as a spin-off, we show that the DML estimator is either degenerate or inconsistent by analyzing its gradient. Then, in Sec. \ref{sec:he}, we present the Hessians of both functions and, finally, we discuss the efficient evaluation of the gradients and Hessians in Sec. \ref{sec:ee}. 

\section{Gradient expressions and inconsistency of the DML estimator}
\label{sec:ge}

For simplicity, we present the gradients of $\FL_D$ and $\FL_C$, denoted $\vg_D$ and $\vg_C$, noting that the gradient of $\FL_S$ is just the sum of the last two, $\vg_S=\vg_D+\vg_C$, due to (\ref{eq:23}).

 $\vg_D$ and $\vg_C$ can be divided into blocks corresponding to the $\vtheta$ and $\vlam$ vectors as follows,
\begin{align}
\label{eq:69}\nonumber
\vg_D\equiv
\begin{bmatrix}
  \vg_{D\theta}\\ \vg_{D\lambda}
\end{bmatrix},
  &&
\vg_C\equiv
\begin{bmatrix}
  \vg_{C\theta}\\ \vg_{C\lambda}
\end{bmatrix}.    
\end{align}
The blocks are the following
{\renewcommand{\arraystretch}{1.5}
\begin{equation}
\label{eq:56}
\begin{array}{r@{}l@{}l}  
\vg_{D\theta}&=&2N\textrm{Re}\{\textrm{diag}\{\mphi^\dagger\mr_{z\lambda}(\mi_M-\
\mpr)\md\}\}\\
\vg_{D\lambda} &=&2N\mlam^{-1}\textrm{diag}\{\SiM-(\mi_M-\mpr)\mr_{z\lambda}(\mi_M-\mpr)\}\\
\vg_{C\theta}&=-&2N\textrm{Re}\{\textrm{diag}\{\mm_{z\lambda}\mphi^H\mr_{z\lambda}(\mi_M-\mpr)\md\}\}\\
\vg_{C\lambda}&=&2N\mlam^{-1}\textrm{Re}\{\textrm{diag}\{\mpr-2\Srzl\mpr_z\}\},
\end{array}
\end{equation}
}
where
\begin{equation}
\label{eq:232}\nonumber
[\md]_{\cdot,k}\equiv \mlam\frac{\partial}{\partial\theta_k}\vphi_o(\theta_k),\;\;k=1,\,\ldots,\,K.
\end{equation}
 The derivation of $\vg_{D\lambda}$ and $\vg_{C\lambda}$ can be found in Ap. \ref{sec:dg} and that of $\vg_{D\theta}$ and $\vg_{C\theta}$ in the complementary material. 

The expressions in (\ref{eq:56}) allow us to check the consistency of the DML and SML estimators. Let us check both simultaneously. If either the DML or SML  estimator is consistent in $\vtheta$ and $\vlam$, then for high $N$ and with high probability we have the following facts,

\begin{enumerate}
\item The data correlation matrix is, approximately,
\begin{equation}
\label{eq:49}
\Srzd\approx \Sphio\mr_s\Sphio+\mlam^{-2},
\end{equation}
where the right hand side is evaluated at the true values of $\vtheta$ and $\vlam$; and where $\mr_s$ is the signal covariance matrix for the SML estimator, or we assume the existence of the asymptotic covariance
\begin{equation}
\label{eq:73}\nonumber
\Srs=\lim_{N\to\infty}\frac{1}{N}\ms\ms^H.
\end{equation}
for the DML estimator. (Note that the number of columns of $\ms$ is $N$.)

\item The consistency assumption implies that the DML or SML estimate is close to the true values of $\vtheta$ and $\vlam$. Therefore, (\ref{eq:49}) also holds if its right-hand side is evaluated at the estimates of $\vtheta$ and $\vlam$ rather than at the true values of these vectors. 

\item The ML estimate corresponds to a critical point of either $\FL_{D}$ or $\FL_{S}$, i.e, we either have
\begin{equation}
\label{eq:16}
\vg_{D\theta}(\vtheta,\vlam)=\vzer,{}\;\;\;\;\vg_{D\lambda}(\vtheta,\vlam)=\vzer
\end{equation}
or
\begin{equation}
\label{eq:119}
\vg_{S\theta}(\vtheta,\vlam)=\vzer,{}\;\;\;\;\vg_{S\lambda}(\vtheta,\vlam)=\vzer.
\end{equation}
\end{enumerate}

Now we can show that these three assertions are incompatible for the DML estimator and compatible for the SML estimator. More precisely, all equations in 3) hold if 1) and 2) are assumed, except for the second one in (\ref{eq:16}), i.e, $\vg_{D\lambda}\neq \vzer$. Let us prove this assertion. First, note that (\ref{eq:49}) implies the approximation 
\begin{equation}
\label{eq:18}\nonumber
 \mr_{z\lambda}=\mlam\mr_z\mlam\approx\mphi\mr_s\mphi^H+\mi_M,
\end{equation}
which in turn implies 
\begin{equation}
\label{eq:76}\nonumber
\mphi^H\mr_{z\lambda}(\mi_M-\mpr)\approx \mzer.
\end{equation}
As a consequence, we deduce  ${\vg_{D\theta}\approx0}$ and  ${\vg_{C\theta}\approx0}$ from the expressions of these gradient blocks in (\ref{eq:56}). And, in turn, we have
\begin{equation}
\label{eq:110}\nonumber
\vg_{S\theta}=\vg_{D\theta}+\vg_{C\theta}\approx \vzer.
\end{equation}
So the first equations of (\ref{eq:16}) and (\ref{eq:119}) hold. 

Second, let us check whether ${\vg_{D\lambda}\approx \vzer}$. For this, operate on the expression for $\vg_{D\lambda}$ in (\ref{eq:56}). We have 
\begin{equation}
\label{eq:93}
\begin{split}
\vg_{D\lambda}& \approx
2N\textrm{diag}\big\{
-\mlam^{-1}\cdot\\
&\strut\;\;(\mi_M-\mpr)(\mi_M+ \mphi\mr_s\mphi^H)(\mi_M-\mpr)+\mlam^{-1}\big\} \\
&=2N\textrm{diag}\{-\mlam^{-1}(\mi_M-\mpr)+\mlam^{-1}\}\\
&=2N\textrm{diag}\{\mlam^{-1}\mpr\}.
\end{split}
\end{equation}
If $\mq$ denotes a matrix whose columns are an ortho-normal basis for the span of $\mphi$, then ${\mpr=\mq\mq^H}$ and the $m$th component of (\ref{eq:93}) can be expressed as
\begin{equation}
\label{eq:50}\nonumber
[\vg_{D\lambda}]_m\approx 2N\frac{\|[\mq]_{\cdot,m}\|^2}{\lambda_m}\geq 0.
\end{equation}
Since $\|[\mq]_{\cdot,m}\|^2$ is positive for at least one index $m$, we have ${\vg_{D\lambda}\neq \vzer}$. This proves that the previous conditions are incompatible for the DML estimator. Therefore, this last estimator is either degenerate or inconsistent. 

And finally, let us check whether $\vg_{S\lambda}\approx \vzer$. For this, we need to consider the expression of $\vg_{C\lambda}$ in (\ref{eq:56}), but let first us prove that $\mpr_z\mr_{z\lambda}\approx\mpr$. We have
\begin{multline}
  \nonumber  
  \mpr_z\mr_{z\lambda}=\mphi(\mphi^H\mr_{z\lambda}\mphi)^{-1}\mphi^H\mr_{z\lambda}\\
  \approx\mphi(\mphi^H(\mi_M+\mphi\mr_s\mphi^H)\mphi)^{-1}\mphi^H(\mi_M+\mphi\mr_s\mphi^H)\\
  =\mphi(\mphi^H\mphi+\mphi^H\mphi\mr_s\mphi^H\mphi)^{-1}\mphi^H(\mi_M+\mphi\mr_s\mphi^H)\\
  =\mphi((\mi_K+\mphi^H\mphi\mr_s)\mphi^H\mphi)^{-1}\mphi^H(\mi_M+\mphi\mr_s\mphi^H)\\
  =\mphi(\mphi^H\mphi)^{-1}(\mi_K+\mphi^H\mphi\mr_s)^{-1}\mphi^H(\mi_M+\mphi\mr_s\mphi^H)\\
  =\mphi(\mphi^H\mphi)^{-1}(\mi_K+\mphi^H\mphi\mr_s)^{-1}(\mphi^H\mphi\mr_s\mphi^H+\mphi^H)\\
  =\mphi(\mphi^H\mphi)^{-1}(\mi_K+\mphi^H\mphi\mr_s)^{-1}(\mi_K+\mphi^H\mphi\mr_s)\mphi^H\\
   =\mphi(\mphi^H\mphi)^{-1}\mphi^H=\mpr.\hfill\strut
\end{multline}
Now, let us operate on the expression of $\vg_{C\lambda}$ in (\ref{eq:56}) assuming this last approximation, recalling that $\Srzl=\mlam\Srzd\mlam$, and using the property $\diag\{\mpr\mlam^{-1}\}=\diag\{\mlam^{-1}\mpr\}$. We have
\begin{equation}
\label{eq:12}\nonumber
\begin{split}
  \vg_{C\lambda}& = 2N(\textrm{Re}\{\textrm{diag}\{\mpr\mlam^{-1}-2\mr_z\mlam\mpr_z\}\})\\
  &= 2N(\textrm{Re}\{\textrm{diag}\{\mpr\mlam^{-1}-2\mlam^{-1}\mr_{z\lambda}\mpr_z\}\})\\
  &\approx 2N(\textrm{Re}\{\textrm{diag}\{\mpr\mlam^{-1}-2\mlam^{-1}\mpr\}\})\\
  &=2N(\textrm{Re}\{\textrm{diag}\{\mlam^{-1}\mpr-2\mlam^{-1}\mpr\}\})\\
  &=-2N\textrm{diag}\{\mlam^{-1}\mpr\}.
\end{split}
\end{equation}
Therefore, from (\ref{eq:93}), we have
\begin{equation}
\label{eq:75}\nonumber
\vg_{S\lambda}=\vg_{D\lambda}+\vg_{C\lambda}\approx \vzer.
\end{equation}
So, we conclude that the three conditions are compatible for the SML estimator.  

\section{Hessian expressions}
\label{sec:he}

It is only necessary to present the Hessians of $\FL_D$ and $\FL_C$, denoted $\mh_D$ and $\mh_C$, given that the Hessian of $\FL_S$ is $\mh_S=\mh_D+\mh_C$ due to (\ref{eq:23}). $\mh_D$ and $\mh_C$ can be divided into blocks corresponding to the $\vtheta$ and $\vlam$ vectors as follows,
\begin{align}
\label{eq:94}\nonumber
\mh_D\equiv
\begin{bmatrix}
  \mh_{D\theta\theta}& \mh_{D\theta\lambda}\\
    \mh_{D\theta\lambda}^T& \mh_{D\lambda\lambda}
\end{bmatrix},\;\;
                          &
\mh_C\equiv                             
\begin{bmatrix}
  \mh_{C\theta\theta}& \mh_{C\theta\lambda}\\
    \mh_{C\theta\lambda}^T& \mh_{C\lambda\lambda}
\end{bmatrix}.
\end{align}
Define first the matrix
\begin{equation}
\label{eq:108}\nonumber
[\md_{2}]_{\cdot,k}\equiv \mlam\frac{\partial^2}{\partial\theta_k^2}\vphi_o(\theta_k),\;k=1,\ldots,\,K.
\end{equation}
The first block is $\mh_{D\theta\theta}$, 
\begin{equation}
\label{eq:233}
{\renewcommand{\arraystretch}{1.5}
\begin{array}{r@{\,}l} 
\mh_{D\theta\theta}
&=2N\Sre\Big\{ \\ &\mm\odot(\md^H
    (\SiM-\mpr)\mr_{z\lambda}(\SiM-\mpr)\md)^T
  \\
  &-
(\Sphi^\dagger \md)\odot(\Sphi^\dagger\mr_{z\lambda}(\SiM-\mpr)\md)^T
        \\
  & -
(\Sphi^\dagger\mr_{z\lambda}(\SiM-\mpr)\md)\odot(\Sphi^\dagger\md)^T 
  \\
  &-
(\Sphi^\dagger\mr_{z\lambda}(\Sphi^\dagger)^H)\odot(\md^H(\SiM-\mpr)\md)^T
    \\
  &+
\SiK\odot(\Sphi^\dagger
   \mr_{z\lambda}(\SiM-\mpr)\md_2)^T
   \Big\}.
\end{array}}
\end{equation}
Note that this block is equal to $\mh_{Do}$ in (\ref{eq:244}) if we set $\mlam=\SiM$. The remaining blocks are the following,
\begin{equation}
  \label{eq:245}
  {\renewcommand{\arraystretch}{1.5}
\begin{array}{r@{\,}l}
\mh_{D\theta\lambda}& =
4N\Sre\big\{
  (\Sphi^\dagger\mr_{z\lambda}(\SiM-\mpr))\odot((\SiM-\mpr)\md)^T\\
&\;\;\;+
(\md^H(\SiM-\mpr)
    \mr_{z\lambda}(\SiM-\mpr))\odot (\Sphi^\dagger)^*
\big\}\mlam^{-1},
\end{array}
}
\end{equation}
\begin{equation}
  \label{eq:246}
  {\renewcommand{\arraystretch}{1.5}
\begin{array}{l}    
  \mh_{D\lambda\lambda}  =2N\mlam^{-1}\Big(\\
                           \re
\Big\{
(4\mpr-\SiM)\odot((\SiM-\mpr)\mr_{z\lambda}(\SiM-\mpr))^T\Big\}\hspace{0.3cm}{}
\\ {}\hfill -\SiM\Big)
\mlam^{-1}, 
\end{array}}
\end{equation}
\begin{equation}
  \label{eq:247}\nonumber
  {\renewcommand{\arraystretch}{1.5}
   \begin{array}{l}
     \mh_{C\theta\theta}
=
      2N\Sre\Big\{ (
      \Smzl\Sphi^H\Srzl(\SiM-\mpr)\md)\odot(\Sphi^\dagger\md)^T\\
     +\mm\odot(\md^H
      (\SiM-\mpr)\md)^T\\
      -\SiK\odot(
      \Smzl\Sphi^H\Srzl(\SiM-\Sp)\md_2)^T\\
      -\Smzl\odot(\md^H(\SiM-\Srzl\Spz)\Srzl(\SiM-\Sp)\md)^T\\
       +(\Smzl\Sphi^H\Srzl\md)\odot
      (\Smzl\mphi^H  \Srzl(\SiM-\Sp)\md)^T\Big\},                                                            
    \end{array}}
\end{equation}
\begin{equation}
\label{eq:248}
{\renewcommand{\arraystretch}{1.5} 
  \begin{array}{r@{\,}l}
\mh&_{C\theta\lambda}=4N\textrm{Re} \Big\{(\md^H (\SiM-\mpr)
             )\odot (\Sphi^\dagger)^*
    \\  
           &(\Smzl\Sphi^H)\odot
             (\Srzl(\SiM-\Spz\Srzl)\md)^T\\
           &(\md^H(\SiM-\Srzl\Spz))\odot
             (\Srzl\Sphi\Smzl)^T \Big\} \mlam^{-1},
    \end{array}}
\end{equation}
\begin{equation}
\label{eq:249}
{\renewcommand{\arraystretch}{1.5}
    \begin{array}{r@{\,}l}
\mh_{C\lambda\lambda} & =
            2N\mlam^{-1}\textrm{Re}\Big\{(\SiM-2\mpr)
            \odot\mpr^T
  \\
& {}\;\;\;  -4(\Srzl(\SiM-\Spz\Srzl))
       \odot
  \Spz^T\\
& {}\;\; \; -2
    (\Srzl\Spz)\odot(\SiM-2\Srzl\Spz)^T\Big\}\mlam^{-1}.      
    \end{array}}
\end{equation}
The derivations of these blocks can be found in the complementary material.

\section{Efficient evaluation of the gradient and Hessian of $\FL_D$ and $\FL_S$}
\label{sec:ee}

The gradient and Hessian expressions in Secs. \ref{sec:ge} and \ref{sec:he} may seem  to involve a large computational burden. However, this is only an initial impression given that there are multiple ways to reduce their complexity, 

\begin{itemize}

\item Given two matrices $\ma$ and $\mb$ of proper size, the expression $\diag(\ma\mb)$ can be more efficiently computed by adding up the columns of  $\ma\odot\mb^T$. This simplifies all the gradient blocks in (\ref{eq:56}).

\item A few Hessian summands of the form ``$(\cdots)\odot \mi_K$'' just involve the computation of the diagonal of a matrix product. For example, the summand $(\mm\mphi^H\md_2)\odot\mi_K$ of $\mh_{C\theta\theta}$ only requires to compute the diagonal components of the left side, i.e, to add up the columns of $(\mm\mphi^H)^T\odot\md_2$.
  
\item Many Hessian terms involve either the product $(\mi_M-\mpr)\Srzl$ or $\Srzl{(\mi_M-\Spz\Srzl)}$, which are small when $\vtheta$ and $\vlam$ are close to their true values, and can usually be neglected in the computation of a Newton iteration. Thus, we may consider evaluating only  the following number of summands for each Hessian block
\begin{equation}
\label{eq:95}\nonumber\nonumber
\begin{array}{c|c|c|c|c|c}
  \mh_{D\theta\theta} & \mh_{D\theta\lambda} & \mh_{D\lambda\lambda} & \mh_{C\theta\theta} & \mh_{C\theta\lambda} & \mh_{C\lambda\lambda} \\ \hline
  1 & 0 & 1 & 1 & 1 & 2
\end{array}
\end{equation}

This simplification is similar to the ones performed in the scoring or Modified Variable Projection (MVP) methods in uniform-noise DOA estimation, \cite{Haykin93,Kaufman75}.

\item There are many repeated matrix products such as $\Sphi^\dagger \Srzl$, ${(\mi-\Sp)}\Sdphi$, and $\Smzl\Sphi^H$ that must be computed only once.

\item The QR decomposition of $\mphi$ simplifies some of the computations. If ${\mphi\mr=\mq}$ with triangular $\mr$ and $\mq^H\mq=\mi_K$, then
\begin{gather}
\label{eq:62}\nonumber
\mm=\mr\mr^H,\;\;\mphi^\dagger=\mr\mq^H,\nonumber \\
\mpr=\mq\mq^H, |\mc|=|\mq^H\Srzl\mq|.\nonumber
\end{gather}
Also, the product $\mpr\ma$ for any matrix $\ma$ is more efficiently computed as $\mq(\mq^H\ma)$.

\item A product of the form ``$(\ldots)\mlam$'' is the same as multiplying each row of the left-side matrix by the corresponding component of $\mlam$. The computation of $\mlam^{-1}$ just involves the inversion of its diagonal components.

\item The Hessian summands have the form $\text{Re}\{\ma\odot\mb\}$ for equal-size complex matrices $\ma$ and $\mb$. But these summands can be obtained with approximately half complexity if the real part is taken first, i.e, rather than $\text{Re}\{\ma\odot\mb\}$ we may compute
\begin{equation}
\label{eq:63}\nonumber
\text{Re}\{\ma\}\odot \text{Re}\{\mb\}-\text{Im}\{\ma\}\odot \text{Im}\{\mb\}.
\end{equation}
\end{itemize}

We have computed polynomials in $M$ and $K$ for the number of arithmetic operations required to evaluate the various cost functions, gradients, and Hessians in this paper. The costs of computing $\FL_D$ and $\FL_S$ are respectively given by  
\begin{equation}
  \label{eq:240}\nonumber
  \begin{array}{l}
    \Fpr(\FL_D;M,K)\equiv -2K^3+8 K^2 M+8 K M^2\hspace{1.2cm}{}\\
    {}\hfill +2 K M+46 M^2+14
\end{array}
\end{equation}
and
\begin{equation}
  \label{eq:241}\nonumber
  \begin{array}{l}
    \Fpr(\FL_S;M,K)\equiv -2 K^3+24 K^2 M-2 K^2\hspace{0.7cm}{}
    \\
    {}\hfill +16 K M^2+2 K M+2 K+64 M^2+18.
\end{array}
\end{equation}
And the costs of computing $\FL_D$ and $\FL_S$ and their corresponding gradients and full Hessians are respectively given by the polynomials
\begin{equation}
  \label{eq:242}\nonumber
  \begin{array}{l}
    \Fpr(\FL_D,\vg_D,\mh_D;M,K)\equiv 8 K^3+72 K^2 M+38 K^2\\
    {}\hfill +40 K M^2 -4 K M+46 M^2+20
\end{array}
\end{equation}
and
\begin{equation}
  \label{eq:243}\nonumber
  \begin{array}{l}
    \Fpr(\FL_S,\vg_S,\mh_S;M,K)=24 K^3+112 K^2 M+80 K^2\\
    {}\hfill +192 K M^2+37 K M+3 K+236 M^2+3 M+32.
\end{array}
\end{equation}

We can see in the following table these computational burdens for the case $M=11$ and $K=3$ that will be assessed in the next section,

\begin{center}
\begin{tabular}{l|r}
  & Flops\\
 \hline  $ \FL_D$ & 9288\\
  $ \FL_S$ & 15946\\
  $\FL_D,\vg_D,\mh_D$& 27660 \\
  $\FL_S,\vg_S,\mh_S$ & 112003
\end{tabular}
\end{center}

\section{Numerical example}
\label{sec:ne}

We have validated the APN method for the $\FL_D$ and $\FL_S$ cost functions numerically in the following scenario:

{\bf Received signals.} There were three received signals and all of them were complex Gaussian processes. As to their spatial correlation, there were two cases,
\begin{itemize}
\item {\it Uncorrelated signals.} The signal covariance matrix was diagonal with
\begin{equation}
\label{eq:99}
\mr_s=\diag([1,0.64,0.25]).
\end{equation}

The signals were deterministic, i.e, the same realization of $\ms$ was used in all Monte Carlo trials.

\item {\it Correlated signals.} The signal covariance matrix had the form
\begin{equation}
\label{eq:100}\nonumber
\mr_s=\mur\diag(\vv)\mur^H,
\end{equation}
where $\mur$ was a realization of a random unitary matrix and
\begin{equation}
\label{eq:101}\nonumber
\vv=[2.337,0.06604,0.0004642]^T.
\end{equation}
\end{itemize}

{\bf Sensor array.} Uniform linear array formed by 11 sensors with half-wavelength spacing. 

{\bf Angles of arrival.} The angles of arrival were the following
\begin{equation}
\label{eq:102}\nonumber
\vtheta=[-0.2513,\,0.1571,\,1.005]^T\;\;\text{(rad)}.
\end{equation}

{\bf Noise inverse deviations.} The inverse deviations $\lambda_m$ followed a linear trend with the sensor index of the form
\begin{equation}
\label{eq:103}\nonumber
\lambda_m=c\,\bigg(1+9\frac{m-1}{M-1}\bigg),\;\;m=1,\,2,\ldots,M, 
\end{equation}
where $c$ was selected in order to ensure a given SNR. Note that from this trend we have $\lambda_M/\lambda_1=10$ and, therefore, the noise power varies 20 dB along the array.

{\bf Estimators.} We tested the following estimators,

\begin{itemize}

\item MUSIC. Multiple Signal Classification estimator.
\item DMLo. DML estimator assuming uniform noise and computed through the APN method.
\item DML. DML estimator of $\vtheta$ and $\vlam$. 
\item DML-alt. DMLo estimator followed by alternate Newton iterations on $\FL_D$ in the $\theta$ and $\vlam$ parameterizations. The initial estimate of $\vlam$ was obtained using the method in \cite{Madurasinghe05}.
\item SML. DMLo followed by Newton maximization of $\FL_S$ using the full Hessian.
\item SML-alt. The same as DML-Alt but with alternate maximization of $\FL_S$.
\item SML-red. The same as SML but neglecting some of the Hessian summands as explained in Sec \ref{sec:he}.

\end{itemize}

{\bf Number of Monte Carlo trials.} We performed 1000 Monte Carlo trials.

\begin{figure}
\includegraphics{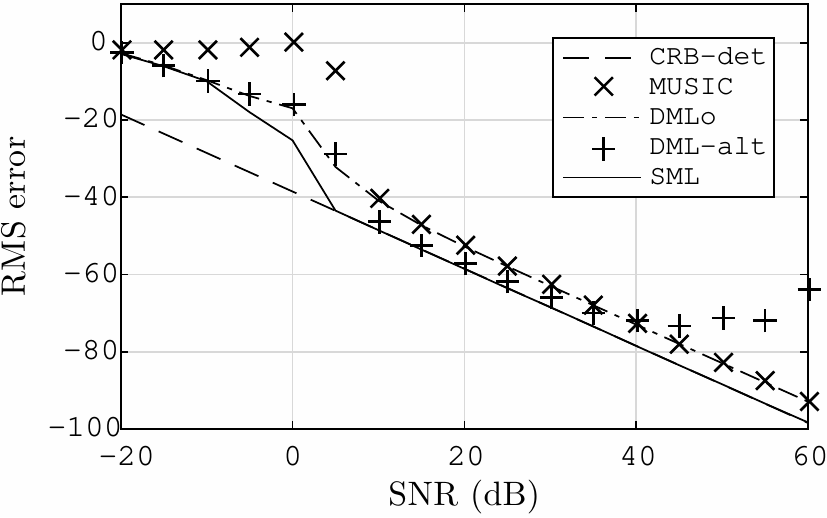}
\caption{\label{fig:2} RMS error performance of MUSIC, DMLo, DML-alt, and SML for  uncorrelated signals.}
\end{figure}

\begin{figure}
\includegraphics{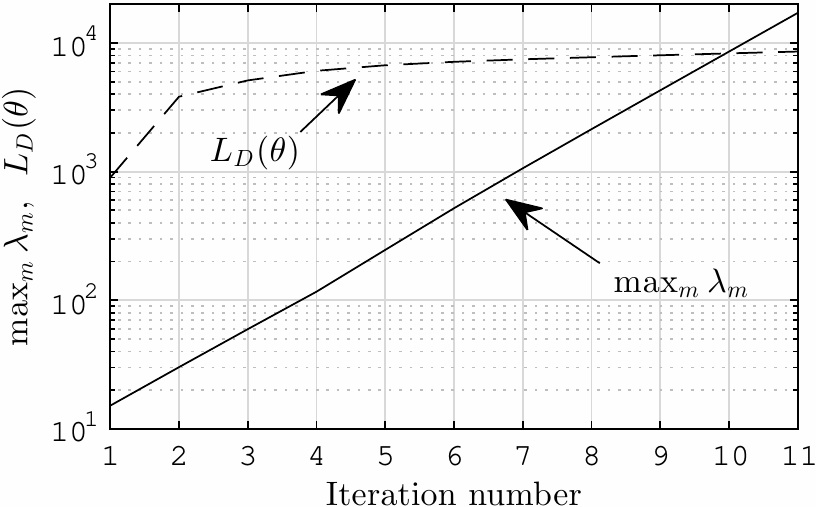}
\caption{\label{fig:1} Iteration number in Newton's method for the DML estimator versus $\max_m \lambda_m$ and $L_{D}(\mathbf{\theta})$.}
\end{figure}

Fig. \ref{fig:2} shows the root-mean-square (RMS) error performance of MUSIC, DMLo, DML-alt, and SML for uncorrelated signals. We can see that MUSIC and DMLo reach an RMS error floor above the CR bound, and this can be attributed to their inability to estimate the sensor noise powers. DML-alt achieves the CR bound at intermediate SNRs but fails to do so at high SNRs. Finally, SML reaches the CR bound at intermediate and high SNRs. In this figure, DML is missing because Newton's method produces a divergent $\vlam$ estimate and, therefore, DML is unavailable. This can be explained by the problem related with DML already commented in Sec. \ref{sec:ge}. Fig. \ref{fig:1} shows this phenomenon for a specific realization, for which Newton's method is initialized with the true values of $\vtheta$ and $\vlam$ and ${\text{SNR}=40}\;\text{dB}$. We can see in this figure that the Newton iteration achieves an increase in the cost function value every time, but $\max_m\lambda_m$ diverges, i.e, at least one sensor noise power is taken as zero approximately. Obviously, this is a degenerate result.

\begin{figure}
\includegraphics{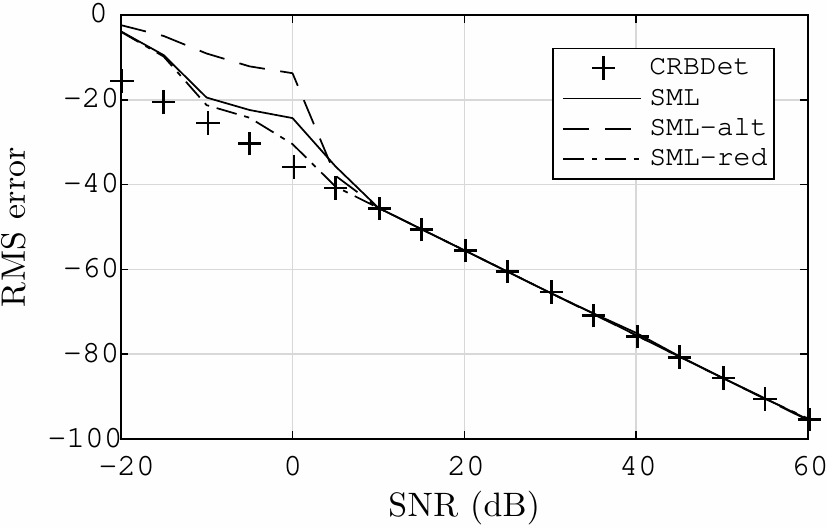}
\caption{\label{fig:3} RMS error of SML estimator computed in three different ways: SML, SML-alt, and SML-red.}
\end{figure}
\begin{figure}
\includegraphics{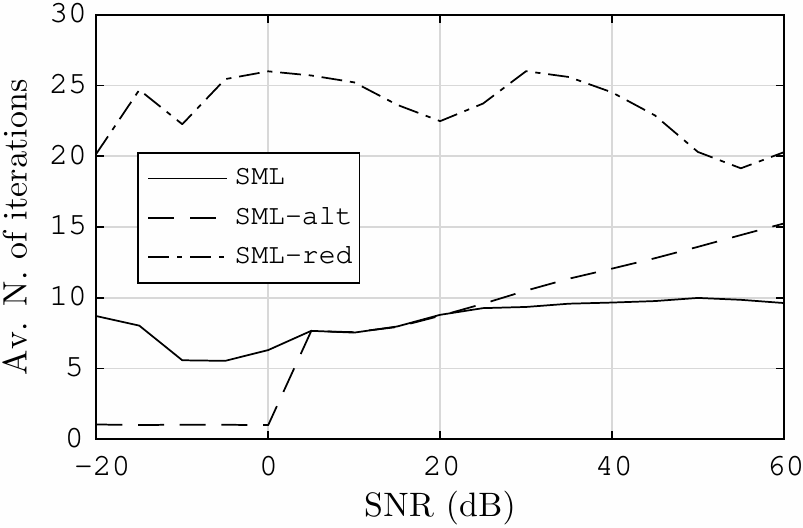}
\caption{\label{fig:4} Average number of iterations of SML, SML-alt, and SML-red.}
\end{figure}

In Fig. \ref{fig:3} we can see the RMS error performance of the SML estimator in the same scenario but computed in three different ways: SML, SML-alt, and SML-red. We can see in this figure that the three computation methods only produce some difference at low SNRs, where SML-alt has the worst performance and SML-red the best. Fig. \ref{fig:4}. shows the average number of iterations for these three methods. 
We can see that SML-alt only requires one iteration when initialized with the DMLo estimate at low SNRs. The problem here is that SML-alt is producing overlapping values of $\theta$, i.e, identical angles, and this occurrence stops the Newton iteration and the value returned by the method is the initial estimate (DMLo estimate). SML-red requires more that twice the number of iterations than SML. However, recall that SML-red only used a fraction of the Hessian terms and, therefore, SML-red iterations are cheaper than SML iterations computationally. 

\begin{figure}
\includegraphics{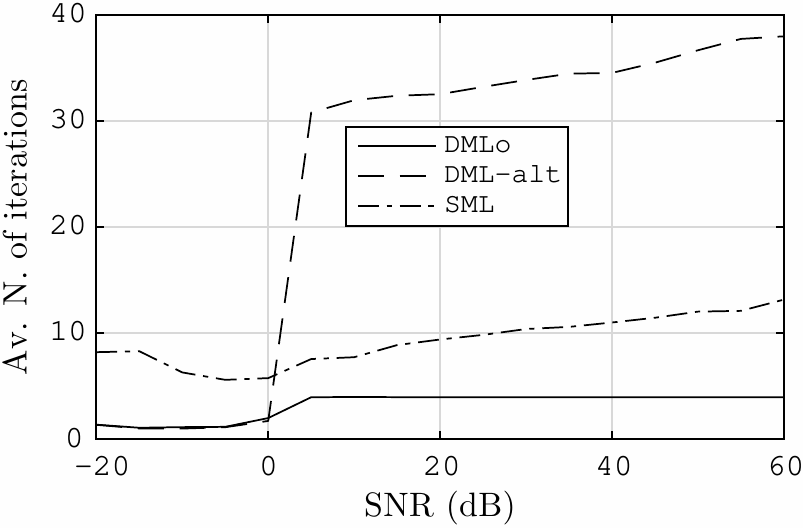}
\caption{\label{fig:7} Number of iterations required by DMLo, DML-alt, and SML in Fig. \ref{fig:1}.}
\end{figure}
Fig. \ref{fig:7} shows the number of iterations required by DMLo, DML-alt, and SML. For DMLo the number of iterations is the one in the last application of Newton's method in (\ref{eq:86}). Note that this number is small for DMLo and SML and high for  DML-alt at most SNRs. Again, we may suspect that the degeneracy or inconsistency of the DML estimator is producing the high number of iterations in DML-alt. Besides, DML-alt is a coordinate ascent method, i.e, the cost function is increased by varying $\vtheta$ and $\vlam$ in turn, and this usually leads to a higher number of iterations. 

\begin{figure}
\includegraphics{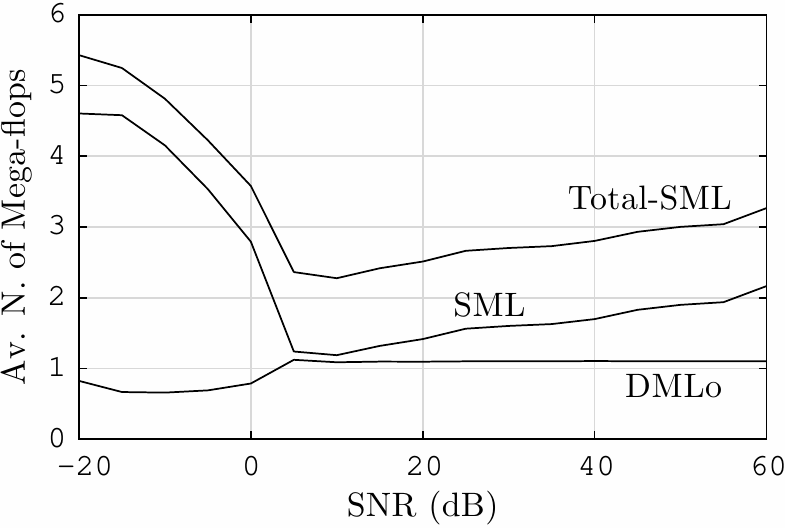}
\caption{\label{fig:8} Average number of mega-flops for DMLo and SML estimators.}
\end{figure}
Fig. \ref{fig:8} shows the computational burden of DMLo and SML, measured in average number of mega-flops. The cost represented by the DMLo curve includes
\begin{itemize}
\item The computation of the Newton iterations on $\FL_{D}$ for $K=1$, 2, and 3. These iterations often required additional computations of $\FL_{D}$ whenever $\mu_{k,r}$ is reduced in (\ref{eq:86}).  
\item The initial line searches for adding a new angle for $K=1$, 2, and 3.
\end{itemize}
The SML curve only stands for the cost of refining the DML estimate for ${K=3}$ through Newton's method applied to $\FL_S$. Finally, the Total-SML curve represents the cost for the whole APN method (sum of DMLo and SML curves), i.e, for computing the SML estimate with $K=3$. Note that this total cost is small at any SNR: from 2.2 to 3.2 Mega-flops. Another feature in Fig. \ref{fig:8} is the increase at low SNRs of the computational burden. This increase is produced by the need to truncate the ascendant direction one or more times by reducing $\mu_{k,r}$ or $\mu_{r}$ in either (\ref{eq:86}) or (\ref{eq:87}).

\begin{figure}
\includegraphics{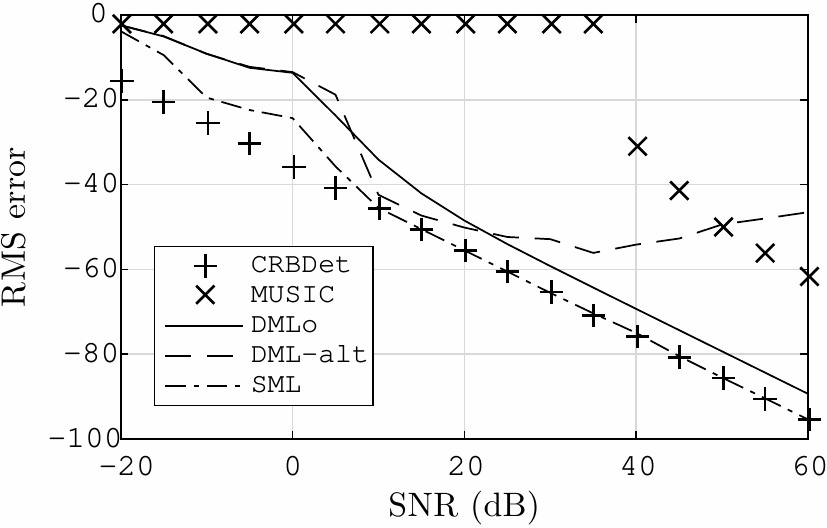}
\caption{\label{fig:5} RMS error performance of MUSIC, DMLo, DML-alt, and SML for correlated signals.}
\end{figure}
\begin{figure}
\includegraphics{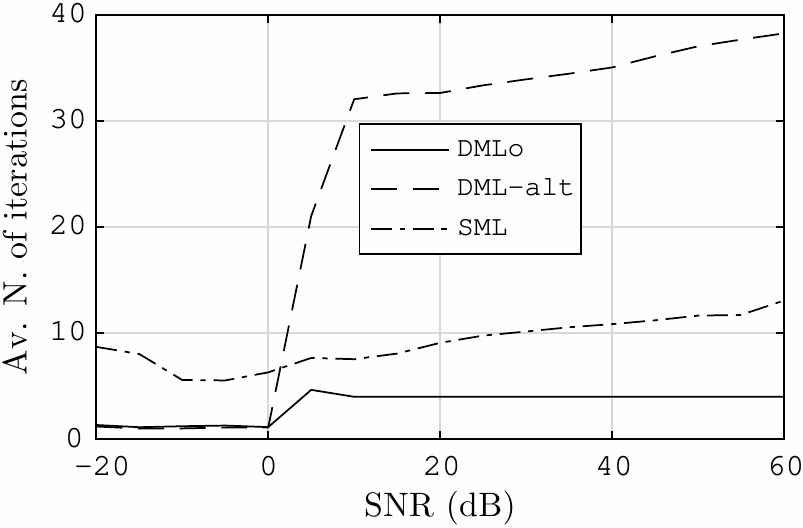}
\caption{\label{fig:6} Average number of iterations of DMLo, DML-alt, and SML.}
\end{figure}

Finally, we can see in Fig. \ref{fig:5} the RMS error performance assuming correlated signals. Note that MUSIC fails as could be expected, and DML-alt fails to reach the CR bound at most SNRs. Again, this last behavior can be explained by the problem related with the DML estimator already discussed in Sec. \ref{sec:ge}. The other estimators perform as in the uncorrelated-signals case in Fig. \ref{fig:2}. Finally, Fig. \ref{fig:6} shows the average number of iterations for correlated signals, and the conclusions that can be drawn are similar to those for Fig. \ref{fig:7}.
\section{Conclusions}

We have presented an efficient method for computing maximum likelihood (ML) estimates of the directions of arrival (DOA) to an array of sensors, assuming unknown sensor noise powers. The method, termed Alternate Projection Newton (APN) method, consists of adding angle estimates sequentially through two steps. In the first, a new angle is added  through a line search and, in the second, the set of angle estimates available is refined using Newton's method. A key part of the APN method is given by closed-form expressions of the ML cost functions (stochastic and deterministic) that allow for the efficient implementation of the Newton iterations. The analysis of the deterministic ML (DML) cost function's gradient reveals an unexpected drawback of the corresponding estimator: it is either degenerate or inconsistent. The APN method is assessed in the paper numerically.

\appendices

\section{Derivation of gradient expressions}
\label{sec:dg}

In the sequel, we let $x$ denote any of the components of $\vlam$ and the sub-script $()_x$ denote differentiation in that variable.  Thus, for instance, if $x$ is $\lambda_m$ then $\mpr_x$ denotes
\begin{equation}
\label{eq:97}\nonumber
\frac{\partial}{\partial \lambda_m}\mpr.
\end
{equation}

We will require the following formulas,

\begin{itemize}
\item We will repeatedly use the fact that the product of two diagonal matrices can be commuted, i.e, $\mlam^{-1}\mlam_x=\mlam_x\mlam^{-1}$.
  
\item For a square invertible matrix $\ma$, Jacobi's formula states that the derivative in a variable $x$ of $\log|\ma|$ is
\begin{equation}
\label{eq:98}
(\log|\ma|)_x=\tr\{\ma^{-1}\ma_x\}.
\end{equation}

\item  The derivatives in $x$ of $\Srzl$ and $\Sp$ can be computed by means of the product derivative rule, and be concisely expressed as
\begin{gather}
  \label{eq:26}
  \mr_{z\lambda,x}=\mlam_x\mlam^{-1}\mr_{z\lambda}+\mr_{z\lambda}\mlam^{-1}\mlam_x,\\
  \label{eq:32}
  \mpr_x=\mpr\mlam^{-1}\mlam_x+\mlam_x\mlam^{-1}\mpr-2\mpr\mlam^{-1}\mlam_x\mpr.
\end{gather}

\item The product $\Spz\Sp_x$ can be concisely written in terms of $\mlam_x$ through orthogonality properties. Specifically, since $\Spz\Sp=\Spz$, we have
\begin{multline}
\label{eq:38}
\Spz\Sp_x=(\Spz\Sp)_x-\Spzx\Sp=\Spzx(\mi-\Sp)\\
=(\Sphi_x\Smzl\Sphi^H+\Sphi\Smzlx\Sphi^H+\Sphi\Smzl\Sphi_x^H)(\mi-\Sp)\\
=\Sphi\Smzl\Sphi_x^H(\mi-\Sp)=\Sphi\Smzl\Sphi^H\Slam^{-1}\Slam_x(\mi-\Sp)\\
=\Spz\Slam^{-1}\Slam_x(\mi-\Sp).\hfill{}
\end{multline}

\end{itemize}

\subsection{Gradient of $\FL_D$ in $\vlam$, $\vg_{D\lambda}$}

Let us derive the expression of $\vg_{D\lambda}$. First, we differentiate (\ref{eq:59}) in $x$ using (\ref{eq:98}),
\begin{multline}
  \nonumber
  \FL_{D,x}=
  N\Big(2\,\textrm{tr}\{\mlam^{-1}\mlam_x\}+\textrm{tr}\{\mpr_x\mr_{z\lambda}\}\\
-\textrm{tr}\{(\mi_M-\mpr)\mr_{z\lambda,x}\}\Big).
  \end{multline}
Second, we substitute (\ref{eq:32}) and (\ref{eq:26}) into this last expression and expand the product with $(\SiM-\mpr)$. The result of these operations is 
\begin{multline}
  \nonumber
\FL_{D,x}=N\Big(2\textrm{tr}\{\mlam^{-1}\mlam_x\}
+\textrm{tr}\{\mpr\mlam^{-1}\mlam_x\mr_{z\lambda}\}+\\
\textrm{tr}\{\mlam_x\mlam^{-1}\mpr\mr_{z\lambda}\}-2\textrm{tr}\{\mpr\mlam^{-1}
\mlam_x\mpr\mr_{z\lambda}\}\\
-\textrm{tr}\{\mlam_x\mlam^{-1}\mr_{z\lambda}\}
-\textrm{tr}\{\mr_{z\lambda}\mlam^{-1}\mlam_x\}+ \\
\textrm{tr}\{\mpr\mlam_x\mlam^{-1}\mr_{z\lambda}\}
+\textrm{tr}\{\mpr\mr_{z\lambda}\mlam^{-1}\mlam_x\}
\Big).
\end{multline}
Third, we rotate the products inside the trace operators in order to get $\mlam_x$ on the right hand side. Besides, we use the property $\mlam_x\mlam^{-1}=\mlam^{-1}\mlam_x$. We obtain
\begin{multline}
  \label{eq:28}
  \FL_{D,x}= 
  2N\Big(\textrm{tr}\{\mlam^{-1}\mlam_x\}+\textrm{tr}\{\mr_{z\lambda}\mpr\mlam^{-1}\mlam_x\}
  \\
+  \textrm{tr}\{\mpr\mr_{z\lambda}\mlam^{-1}\mlam_x\}
   -\textrm{tr}\{\mpr\mr_{z\lambda}\mpr\mlam^{-1}\mlam_x\}\hfill \\
  {}\hspace{3cm} -\textrm{tr}\{\mr_{z\lambda}\mlam^{-1}\mlam_x\}\Big)  =\\
  2N\Big(\textrm{tr}\{\mlam^{-1}\mlam_x\}-\textrm{tr}\{
(\mi-\mpr)\mr_{z\lambda}(\mi-\mpr)
\mlam^{-1}\mlam_x
  \}\Big).
\end{multline}
If $x$ is one of the components of $\vlam$, say $\lambda_m$, then $\mlam_x =\vdel_{M,m} \vdel_{M,m}^T$, and we have $\tr\{\ma\mlam_x\}=[\ma]_{m,m}$ for any matrix $\ma$ of proper size. So, to obtain the gradient, we just need to replace $\tr\{\ma\mlam_x\}$ with $\diag\{\ma\}$ in (\ref{eq:28}) for every possible $\ma$. The result of this operation is
\begin{equation}
{\renewcommand{\arraystretch}{1.5}
\label{eq:230}\nonumber
\begin{array}{r@{}l@{}l}
  \vg_{D\lambda} &=&2N\textrm{diag}\{-\mlam^{-1}(\mi_M-\mpr)\mr_{z\lambda}(\mi_M-\mpr)+\mlam^{-1}\}\\
&=&  2N\mlam^{-1}\textrm{diag}\{\SiM-(\mi_M-\mpr)\mr_{z\lambda}(\mi_M-\mpr)\},
\end{array}
}
\end{equation}
which is the second formula in (\ref{eq:56}).

\subsection{Gradient of $\FL_C$ in $\vlam$, $\vg_{C\lambda}$}

First, differentiate (\ref{eq:23}) in $x$ using (\ref{eq:98}), (\ref{eq:109}), and (\ref{eq:125}):
\begin{multline}
  \label{eq:30}
  \FL_{C,x}=(-N\log|\mc|)_x=-N\textrm{tr}\{\mc^{-1}\mc_x\}=\\
  -N\textrm{tr}\{(\mi_M-\mpr+\mpr_z)\{\mi_M-\mpr+\mpr\mr_{z\lambda}\mpr\}_x\}=\\
  -N\textrm{tr}\{(\mi_M-\mpr+\mpr_z)\{-\mpr+\mpr\mr_{z\lambda}\mpr\}_x\}.
\end{multline}
From (\ref{eq:32}), it can be easily checked that $\{-\mpr+\mpr\mr_{z\lambda}\mpr\}_x$ is equal to a sum of terms whose row or column span lies in the span of $\mphi$. This implies $\textrm{tr}\{(\mi_M-\mpr) \{-\mpr+\mpr \mr_{z\lambda} \mpr\}_x\}=\vzer$ and, therefore, (\ref{eq:30}) simplifies to
\begin{equation}
\label{eq:33}\nonumber
\FL_{C,x}= -N\textrm{tr}\{\mpr_z\{-\mpr+\mpr\mr_{z\lambda}\mpr\}_x\}.
\end{equation}
Applying the product derivative rule, we have
\begin{multline}
  \nonumber
\FL_{C,x}=-N(-\textrm{tr}\{\mpr_z\mpr_x\}
+\textrm{tr}\{\mpr_z\mpr_x\mr_{z\lambda}\mpr\} \\
+\textrm{tr}\{\mpr_z\mpr\mr_{z\lambda,x}\mpr\}+\textrm{tr}\{\mpr_z\mpr\mr_{z\lambda}\mpr_x\}).
\end{multline}
Next, we rotate the trace arguments, leaving the derivatives on the right, and apply the property
\begin{equation}
\label{eq:43}
\Spz\Sp=\Sp\Spz=\Spz.
\end{equation}
We obtain
\begin{multline}
\label{eq:35}
\FL_{C,x}=-N(-\textrm{tr}\{\mpr_z\mpr_x\}+\textrm{tr}\{\mr_{z\lambda}\mpr_z\mpr_x\}\\
+\textrm{tr}\{\mpr_z\mr_{z\lambda,x}\})+\textrm{tr}\{\mpr_z\mr_{z\lambda}\mpr_x\}.
\end{multline}
At this point, the fact that $\Sp$, $\Spz$, and $\Srzl$ are Hermitian implies that $-\textrm{tr}\{\mpr_z\mpr_x\}$ and $\tr\{\Spz\Srzlx\}$ are real and $\textrm{tr}\{\mr_{z\lambda}\mpr_z\mpr_x\}^*=\textrm{tr}\{\mpr_z\mr_{z\lambda}\mpr_x\}$. Using these two properties, we may write (\ref{eq:35}) more concisely as
\begin{multline}
  \nonumber
\FL_{C,x}=-N\Sre\big\{-\tr\{\Spz\Spx\}\}+2\tr\{\Srzl\Spz\Spx\}\}+\\
\tr\{\Spz\Srzlx\}\big\}.
\end{multline}
Now, the orthogonality properties and (\ref{eq:38}) imply $\tr\{\Spz\allowbreak\Spx\}=0$. So, we have
\begin{equation}
\label{eq:42}\nonumber
\FL_{C,x}=-N\Sre\big\{2\tr\{\Srzl\Spz\Spx\}\}+\\
\tr\{\Spz\Srzlx\}\big\}.
\end{equation}
Next, we insert the formulas for $\Spx$ and $\Srzlx$ in (\ref{eq:26}) and (\ref{eq:32}),
\begin{multline}
  \nonumber
\FL_{C,x}=-N\textrm{Re}\big\{
2\textrm{tr}\big\{\mr_{z\lambda}\mpr_z(\mpr\mlam^{-1}\mlam_x+\mlam^{-1}\mlam_x\mpr\\
-2\mpr\mlam^{-1}\mlam_x\mpr)\big\}+
\textrm{tr}\big\{\mpr_z(\mr_{z\lambda}\mlam^{-1}\mlam_x+
\mlam^{-1}\mlam_x\mr_{z\lambda})\big\}\big\}.
\end{multline}
This expression can be readily expanded into a sum of trace terms. Then, rotating the trace arguments so that $\Slam_x$ appears on the right-hand side, applying (\ref{eq:43}) and noting that $\tr\{\Srzl\Spz\Slam^{-1}\Slam_x\}^*=\tr\{\Spz\Srzl\Slam^{-1}\Slam_x\}$, we obtain
\begin{equation}
  \label{eq:44}\nonumber
{\renewcommand{\arraystretch}{1.3}
\begin{array}{l}
\FL_{C,x}=2N\textrm{Re}\big\{\textrm{tr}\big\{\mpr\mr_{z\lambda}\mpr_z\mlam^{-1}
\mlam_x\big\} \\ {}\hfill -2\textrm{tr}\big\{\mr_{z\lambda}
 \mpr_z\mlam^{-1}\mlam_x\big\} \big\}
\\
 =2N\textrm{Re}\big\{\textrm{tr}\big\{\mpr\mr_{z\lambda}\mpr_z\mlam^{-1}
\mlam_x\big\}  -2\textrm{tr}\big\{\mlam^{-1}\mr_{z\lambda}
\mpr_z\mlam_x\big\}
\big\}
\\
=2N\textrm{Re}\big\{\textrm{tr}\big\{\mpr\mr_{z\lambda}\mpr_z\mlam^{-1}
\mlam_x\big\}  -2\textrm{tr}\big\{\mr_{z}\mlam
\mpr_z\mlam_x\big\}
\big\}.
\end{array}
}
\end{equation}
Finally, noting that $\mpr\mr_{z\lambda}\mpr_z=\mpr$, we obtain
\begin{multline}
  \nonumber
\FL_{C,x}=2N\textrm{Re}\big\{\textrm{tr}\big\{\mpr\mlam^{-1}
\mlam_x\big\}  -2\textrm{tr}\big\{\mr_{z}\mlam
\mpr_z\mlam_x\big\}
\big\}.
\end{multline}
If we let $x$ run through the variables in $\vlam$ in the same way as we did for (\ref{eq:28}), the result is
{\renewcommand{\arraystretch}{1.5}
\begin{equation}
\label{eq:231}\nonumber
\begin{array}{r@{}l@{}l}  
  \vg_{C\lambda}&=&2N\textrm{Re}\{\textrm{diag}\{\mpr\mlam^{-1}-2\mr_z\mlam\mpr_z\}\}\\
  &=&2N\mlam^{-1}\textrm{Re}\{\textrm{diag}\{\mpr-2\Srzl\mpr_z\}\},
\end{array}
\end{equation}
}
where we have used $\mr_z\mlam=\mlam^{-1}\Srzl$. This is the formula for $\vg_{C\lambda}$ in (\ref{eq:56}).


\bibliographystyle{IEEEbib}

\bibliography{c:/JesusSelva/Data2/Utilities/LaTeX/Bibliography}

\end{document}